\documentclass[useAMS,usenatbib,usegraphicx]{mn2e}
\usepackage{times}
\usepackage{subfigure}
\usepackage{float}
\usepackage{graphicx}
\usepackage{pdfpages}
\usepackage{adjustbox}
\usepackage{amsmath}
\usepackage{amssymb}
\usepackage{booktabs}

\newcommand{\ms}{M$_{\odot}$}
\newcommand{\rs}{R$_{\odot}$}
\newcommand{\ls}{L$_{\odot}$}

\newcommand{\virg}{``}

\title[Post in-spiral common envelope ejecta I]{Properties of the post in-spiral common envelope ejecta I: dynamical and thermal evolution}
\author[Iaconi et al.]{Roberto Iaconi $^{1,2,3}$\thanks{email: roberto.iaconi@kusastro.kyoto-u.ac.jp} \thanks{JSPS International Research Fellow (Graduate School of Science, Kyoto University)}, Keiichi Maeda$^{1}$, Orsola De Marco $^{2,3}$, Takaya Nozawa $^{4}$ \newauthor
and Thomas Reichardt $^{2,3}$\\
$^{1}$Department of Astronomy, Kyoto University, Kitashirakawa-Oiwake-cho, Sakyo-ku, Kyoto 606-8502, Japan\\
$^{2}$Department of Physics \& Astronomy, Macquarie University, Sydney, NSW 2109, Australia\\
$^{3}$Astronomy, Astrophysics and Astrophotonics Research Centre, Macquarie University, Sydney, NSW 2109, Australia\\
$^{4}$Division of Science, National Astronomical Observatory of Japan, Mitaka, Tokyo 181-8588, Japan\\}

\begin{document}

\date{Accepted by MNRAS, \today}

\pagerange{\pageref{firstpage}--\pageref{lastpage}} \pubyear{\the\year{}}

\maketitle

\label{firstpage}

\begin{abstract}
We investigate the common envelope binary interaction, that leads to the formation of compact binaries, such as the progenitors of Type Ia supernovae or of mergers that emit detectable gravitational waves. 
In this work we diverge from the classic numerical approach that models the dynamic in-spiral. We focus instead on the asymptotic behaviour of the common envelope expansion after the dynamic in-spiral terminates. We use the SPH code {\sc phantom} to simulate one of the setups from Passy et al., with a 0.88~\ms, 83~\rs \ RGB primary and a 0.6~\ms \ companion, then we follow the ejecta expansion for $\simeq 50$~yr. Additionally, we utilise a tabulated equation of state including the envelope recombination energy in the simulation (Reichardt et al.), achieving a full unbinding.
We show that, as time passes, the envelope's radial velocities dominate over the tangential ones, hence allowing us to apply an homologous expansion kinematic model to the ejecta. The external layers of the envelope become homologous as soon as they are ejected, but it takes $\simeq 5000$~days ($\simeq 14$~yr) for the bulk of the unbound gas to achieve the homologously expanding regime. We observe that the complex distribution generated by the dynamic in-spiral evolves into a more ordered, shell-like shaped one in the asymptotic regime.
We show that the thermodynamics of the expanding envelope are in very good agreement with those expected for an adiabatically expanding sphere under the homologous condition and give a prediction for the location and temperature of the photosphere assuming dust to be the main source of opacity.
This techniques ploughs the way to determining the long term light behaviour of common envelope transients.
\end{abstract}

\begin{keywords}
stars: AGB and post-AGB - stars: evolution - binaries: close - hydrodynamics - methods: numerical -  dust, extinction
\end{keywords}


\section{Introduction}
\label{sec:introduction}
The common envelope interaction, hereafter CE (\citealt{Paczynski1976}; see also  \citealt{Ivanova2013} for a review), represents a short evolutionary phase in the life of binary systems that takes place when one of the components evolves to the giant stage whilst the other is a much smaller object, such as a main sequence star or a white dwarf. Compact binary white dwarfs, neutron stars and black holes have likely gone through one or more CE events during their evolution. These systems can merge at a later time, potentially generating Type Ia supernovae (\citealt{Webbink1984}), gamma ray bursts (\citealt{Fryer2007}) or emitting detectable gravitational waves \citep{Abbott2016}.

The CE interaction is characterised by the in-spiral of the two stars towards each other in a time-scale comparable to the dynamical time of the giant star ($\lesssim1$~yr), therefore it is usually named dynamic in-spiral. The dynamic in-spiral is triggered by orbital instabilities, such as the Darwin instability (\citealt{Darwin1879}). This leads to a phase of Roche lobe overflow that is unstable and eventually results in the dynamical in-spiral, when energy and angular momentum from the orbit are transferred to the primary's envelope. The physical mechanism regulating the energy exchange is primarily the gravitational drag affecting the gas in the proximity of the companion star(for studies of the gravitational drag in simulations of the CE interaction see \citealt{Ricker2012}, \citealt{Passy2012}, \citealt{Staff2016b}, \citealt{Iaconi2017}, \citealt{MacLeod2017b} and \citealt{Reichardt2019}, while for an analytical study on the phenomenon see \citealt{Ostriker1999}).
The envelope is, as a result, lifted from the potential well of the binary and partly unbound, while the orbital separation between the primary core and the companion is dramatically reduced. 
Orbital energy is unlikely to be the only source leading to the unbinding of the envelope. Additional sources of energy may contribute and probably the most discussed of them is hydrogen and helium recombination (\citealt{Nandez2015}, \citealt{Nandez2016}, \citealt{Ivanova2016}). The efficiency of such energy source has been recently debated, with \citet{Grichener2018}, \citet{Soker2018} and \citet{Wilson2019} arguing that the hydrogen recombination energy might be partly radiated away and therefore not available to do work on the envelope, while \citet{Ivanova2018} argue for a much higher efficiency. Other scenarios that could aid envelope unbinding are envelope fall-back (\citealt{Kuruwita2016}), stellar pulsations (\citealt{Clayton2017}), jets from the companion (\citealt{Shiber2017}, \citealt{Shiber2017b} and \citealt{Shiber2019}), dust-driven winds (\citealt{Glanz2018}) and convection (\citealt{Wilson2019}).

Given the inherent complexity of the dynamic in-spiral phase, its treatment requires 3D hydrodynamic simulations. The first simulations of the common envelope interaction go back more than three decades, though the number has increased dramatically within the last decade. Notable works include \citet{Livio1988}, \citet{Terman1994}, \citet{Rasio1996}, \citet{Sandquist1998}, \citet{Sandquist2000}, \citet{Ricker2008}, \citet{Ricker2012}, \citet{Passy2012}, \citet{Nandez2015}, \citet{Nandez2016}, \citet{Ohlmann2016}, \citet{Ohlmann2016b}, \citet{Iaconi2017}, \citet{Iaconi2018}, \citet{Reichardt2019}, \citet{Chamandy2018}, \citet{Chamandy2019} and \citet{Prust2019}.

The numerical works cited above focus on the physics and the immediate outcomes of the dynamical in-spiral and the mass transfer phase immediately preceding it. Only those simulations including the effects of recombination energy of the gas \citep{Nandez2015,Nandez2016} succeed in unbinding the envelope and only in lower mass stars (\citealt{Nandez2016}, \citealt{Iaconi2018} and Reichardt et al. in preparation). Whether or not recombination energy can be an efficient way to remove the envelope, alone it is not the solution to unbinding the common envelope. 
Recently, a different picture is emerging: the envelope is unbound during an extended phase that follows the dynamical in-spiral. This phase cannot be modelled by current 3D simulations, unless some simplifying assumptions are made. 

One approach to the modelling of timescales much longer than the dynamical timescale is with 1D codes. However, as shown by \citet{Ivanova2016}, reproducing 1D common envelope-like structures by using stellar evolution codes to deposit energy into a giant star is challenging and produces results not particularly similar to those obtained from 3D simulations. This caveat aside, 1D simulations of the common envelope interaction, including the longer thermal timescales have been carried out by, e.g., \citet{Clayton2017} and \citet{Fragos2019}.

Here we take another approach. We postulate that as the unbound envelope expands after all orbital and recombination energy is injected into it, its thermodynamic quantities gradually redistribute, so that the expansion can be described by an adiabatic, homologous kinematic model. Further motivation is provided by the fact that a homologous expansion model can reproduce relatively well the observed ejecta properties of red novae (e.g., the object Vg4332; \citealt{Kaminski2018}), which are possible observational counterparts of CE events (\citealt{Ivanova2013b}).
This reduces the post dynamic in-spiral expansion to a simple analytical model, which can be easily evolved using initial conditions from 3D hydrodynamic simulations. This then allows us to focus on the dynamic, thermodynamic and optical properties of the post-in-spiral CE ejecta over longer time-scales (up to $\simeq 50$~yr). Moreover, if the homologous condition holds this method also allows the study of the ejecta on the time-scales for the formation of planetary nebulae (tens to hundreds thousands of years). 

The calculation of the location and temperature of the photosphere as the envelope expands, accounting for radiation transport, is also greatly simplified in such model and offers many advantages compared to the approach of \citet{Galaviz2017}.

This paper is structured as follows. 
In Section~\ref{sec:homologous_expansion} we introduce the kinematic law of the homologous expansion model and the time dependence of the various thermodynamic quantities if the homologous law is applied to an adiabatic expanding ideal gas. 
In Section~\ref{sec:numerical_choices_and_model_sanity_checks} we discuss our numerical choices and check whether the simulation satisfies the basic conditions of homologous expansion.
In Section~\ref{sec:homologous_expansion_in_CE_ejecta} we test how the simulation reproduces the analytical model during its evolution, considering both the evolution of the dynamic and thermodynamic quantities.
In Section~\ref{sec:determination_of_the_photosphere} we calculate the location and temperature of the photosphere. Finally, we summarise our findings and state our conclusions in Section~\ref{sec:summary_and_conclusions}.


\section{Homologous expansion}
\label{sec:homologous_expansion}
Homologous expansion kinematic models are widely used to model radiation transfer in supernova ejecta simulations (see e.g., \citealt{Ropke2005}, \citealt{Maeda2014}). The assumption of homologous expansion provides the codes with a way to determine the kinematics of the ejecta so that most of the computational power can be focussed in solving the radiation transfer equations.
This model holds under the assumptions that the material is expanding radially and that it has received an energy input that triggered the expansion itself. After that, the amount of energy injected into and lost by the expanding material is assumed to be negligible (i.e., adiabatic conditions). If these conditions are satisfied, then
\begin{equation}
 v_{\mathrm{rad}} = \frac{r}{t - t_0} = \frac{r}{t_{\rm h}} \ ,
 \label{eq:homologous_expansion}
\end{equation}
where $v_{\mathrm{rad}}$ is the radial velocity, $r$ is the radial location of a gas parcel, $t$ is the current time and $t_0$ is time at which homologous expansion is assumed to start. We then define $t_{\rm h} = t - t_0$ as the ``homologous time''.
The starting time of homologous expansion, $t_0$, can be somewhat arbitrary and we will discuss our choice in Section~\ref{sec:homologous_expansion_in_CE_ejecta}.
By differentiating the previous expression it is possible to obtain the radial location $r$ at time $t_{\rm h}$ for the material located at any initial radius $r_{\rm i}$ and homologous time $t_{\rm h,i}$ as
\begin{equation}
 r = \frac{t_{\rm h}}{t_{\rm h,i}} r_{\rm i} \ .
 \label{eq:homologous_expansion_factor}
\end{equation}
Therefore the radius at a certain time $t_{\rm h}$ is determined by the factor $t_{\rm h} / t_{\rm h,i}$.

Assuming that our system is composed of an ideal polytropic gas, for a sphere with a uniform temperature and energy, the thermal energy (non-specific) is directly proportional to the temperature, according to the equation
\begin{equation}
 E_{\textrm{therm}} = \frac{3}{2} n k_{\mathrm{B}} T
 \label{eq:ideal_gas_etherm}
\end{equation}
and the equation of state governing the thermodynamic variables is
\begin{equation}
 P = \rho \frac{E_{\textrm{therm}}}{M} (\gamma - 1) \ ,
 \label{eq:ideal_gas_density}
\end{equation}
where $n$ is the number of particles in the system, $k_{\mathrm{B}}$ is the Boltzman constant, $M$ is the mass of the system, $T$ is the temperature, $P$ is the pressure, $\rho$ is the density and $\gamma$ is the adiabatic index. 
If we couple this with the assumption that the expansion is spherical (trivially correct in the homologous expansion case; $V \propto r^3$, where $V$ is the volume of the sphere) and that the expansion is adiabatic 
(i.e., $dE_{\textrm{therm}}+PdV=0$) with $\gamma = \frac{5}{3}$, for an homologous expanding system it is possible to derive the proportionality laws
\begin{equation}
 E_{\textrm{therm}} \propto t_{\rm h}^{-2} \ ,
 \label{eq:homologous_thermal_energy}
\end{equation}
\begin{equation}
 T \propto t_{\rm h}^{-2} \ ,
 \label{eq:homologous_temperature}
\end{equation}
\begin{equation}
 P \propto t_{\rm h}^{-5} \ ,
 \label{eq:homologous_pressure}
\end{equation}
\begin{equation}
 \rho \propto t_{\rm h}^{-3} \ ,
 \label{eq:homologous_density}
\end{equation}
and
\begin{equation}
 S = \rm{const.} \ ,
 \label{eq:homologous_entropy}
\end{equation}
where $S$ is the entropy of the gas.
Based on Equations~\ref{eq:homologous_thermal_energy} to \ref{eq:homologous_entropy} one can predict the behaviour of the main thermodynamic properties of an adiabatically expanding sphere of ideal gas following the homologous kinematic.


\section{Numerical choices and model sanity checks}
\label{sec:numerical_choices_and_model_sanity_checks}
It is easy to see how the homologous kinematic model described in Section~\ref{sec:homologous_expansion} can be used to describe supernova ejecta, given the very short energy injection timescale and the high radial velocity of the ejecta. 

Here we apply the homologous expansion model to the unbound portion of the CE ejecta obtained from numerical simulations and assess the fidelity of this kinematic recipe on both short and longer time-scales.
However, CE interactions present some substantial differences compared to supernova explosions. Namely, the injection of energy cannot be considered instantaneous and at least initially, the velocity field generated by the dynamic in-spiral is not dominated by the radial component of the velocities (i.e., the tangential components are not negligible). However, as we will show below, it is still possible to approximate the post in-spiral CE ejecta by a homologous model.  

\subsection{Grid vs. SPH codes}
\label{ssec:grid_vs_sph_codes}
To test the model proposed in this work, we have used a CE SPH simulation along the lines of those performed by \citet{Iaconi2017} and \citet{Reichardt2019} using the Smooth Particle Hydrodynamic (SPH) code {\sc phantom} \citep{Price2018}. 

SPH has the ability to follow the ejecta out to large radii from the central binary, while grid simulations have a relatively small computational domains. SPH therefore allows us to compare both the initial and the asymptotic behaviours of the expanding layers of the envelope with our kinematic prescription. Increasing the domain size of a {\it grid} CE simulation so as to follow the ejecta, is not advisable because of the effect of the ``hot vacuum", a numerical expedient used in grid simulation to stabilise the initial star. As discussed by \citet{Iaconi2018} this expedient can have a dynamical effect on the simulation for large domain sizes. Last, the ``hot vacuum" heats the outer layers of the ejecta, altering the thermodynamics properties of the photosphere \citep{Galaviz2017}.

Reichardt et al. (in preparation) have upgraded  {\sc phantom} to include recombination energy by using a tabulated equation of state, similarly to the implementation of \citet{Nandez2015} and \citet{Nandez2016}. 
In this implementation the entire recombination energy payload is thermalised instantly and available to do work. With the parameters of our simulation, the entire envelope is unbound.

\subsection{Numerical setup}
\label{ssec:numerical_setup}
Similarly to \citet{Passy2012} and \citet{Iaconi2017}, we have simulated a red giant branch primary with a total mass of  0.88~\ms\ and a radius of $\simeq 83$~\rs. The stellar core is simulated by a point mass particle with a mass of $\simeq 0.39$~\ms. The main-sequence/WD companion is simulated by a point mass particle of 0.6~\ms \ and it is placed on the primary's surface in circular Keplerian orbit at the beginning of the simulation, something that triggers immediate in-spiral.  The point masses interact with the gas only gravitationally and their gravity is smoothed by a 3~\rs\ softening length (see \citealt{Price2018} for more information on the smoothing procedure).

We use $\simeq 80\,000$ SPH particles. This is the same as the lowest resolution simulations carried out with identical parameters by \citet{Reichardt2019} and was deemed sufficient by their resolution studies for the analysis of the in-spiral and ejecta parameters. Given the longer timescales  considered in this work, increasing the resolution further would have been prohibitive. The simulation conserves energy at the 0.1 percent level and angular momentum at the 0.03 percent level.

We carried out our CE simulation using a tabulated EoS, something that includes the effects of recombination energy. Full details on the treatment of the recombination energy will be presented by  Reichardt at al. (in preparation), but the implementation is identical to that of \citet{Nandez2015} and \citet{Nandez2016}.

The binary systems are simulated from their initial configuration through the dynamic in-spiral phase, and then they are further evolved up to 18\,250~days ($\simeq 50$~yr) after the dynamic in-spiral terminates.

\subsection{Energy injection time-scales}
\label{ssec:energy_injection_timescales}
In order to apply our homologous expansion approximation we need to assess the time at which the CE energy sources have been fully injected. The first source of energy is the orbital energy. This is fully injected by the end of the in-spiral, estimated using the criterion of \citet[][see also \citealt{Passy2012} and \citealt{Iaconi2017}]{Sandquist1998}. This happens at 300 days after the beginning of the simulation. At the end of the dynamic in-spiral the core-companion separation is $\simeq 20$~\rs.

The second source of energy is the recombination energy. The entire envelope becomes unbound by $\simeq 500$~days after the start of the simulation. However, gas is still recombining at that time. By $\simeq 620$~days no more recombination takes place and the energy of the unbound gas remains constant. At the end of the recombination energy injection phase the core-companion separation is $\simeq 17$~\rs.

\subsection{Bound vs. unbound ejecta}
\label{ssec:bound_vs_unbound_ejecta}
Our simulation unbinds 94 percent of the envelope. This value has been calculated by deeming a parcel of gas to be unbound if the sum of its potential, kinetic and thermal energies is positive. In this simulation, all the SPH particles tend to the same asymptotic behaviour. In this work we will not argue whether all the recombination energy should be allowed to do work, or whether there should be other sources of energy at play. By using the gas distribution of our recombination energy simulation we are implicitly assuming that the envelope is (almost) fully ejected because of the orbital energy injection and a second source of energy that operates during or shortly after the dynamical in-spiral. We therefore make the tacit assumption that the kinematic and thermal parameters of the gas are those that result from such assumptions.

\subsection{Radial and tangential velocities}
\label{ssec:radial_velocities_distributions}
\begin{figure*}
\centering     
\includegraphics[scale=0.30, trim=2.5cm 1.0cm 0.0cm 0.0cm]{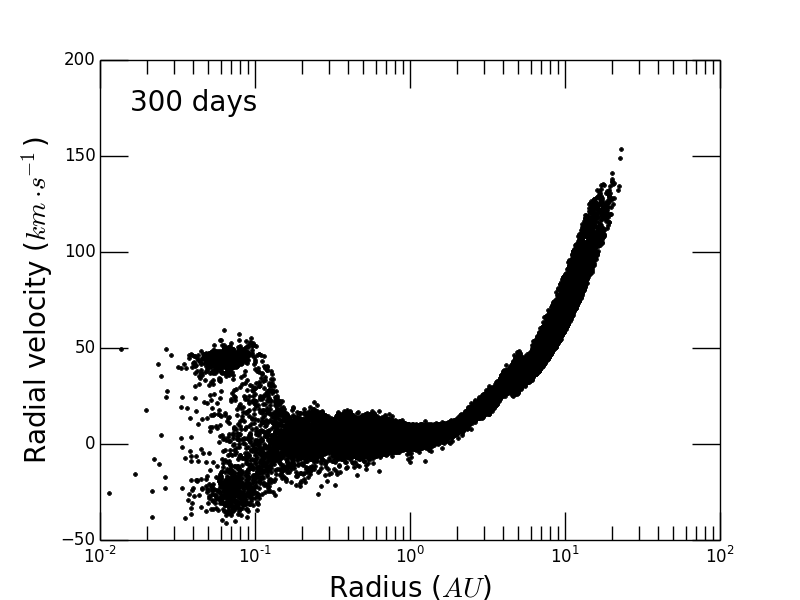}
\includegraphics[scale=0.30, trim=0.0cm 1.0cm 0.0cm 0.0cm]{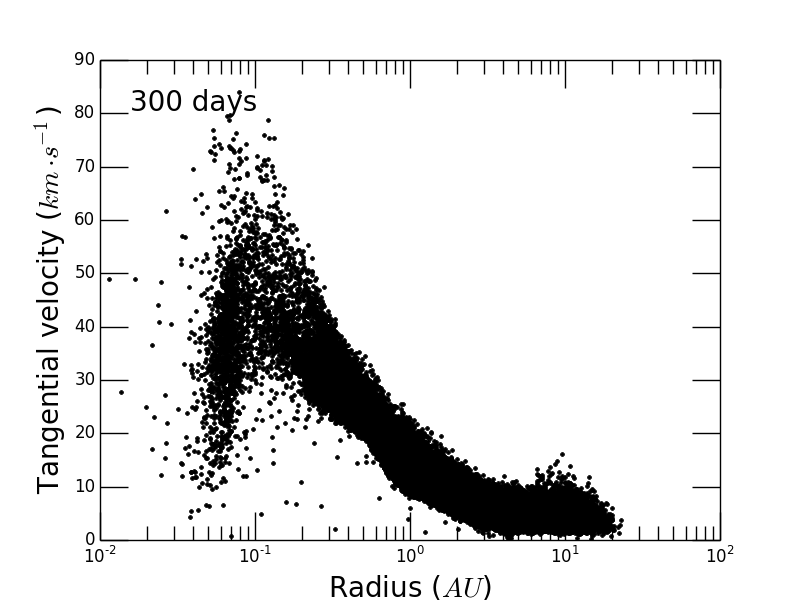}
\includegraphics[scale=0.30, trim=0.0cm 1.0cm 0.0cm 0.0cm]{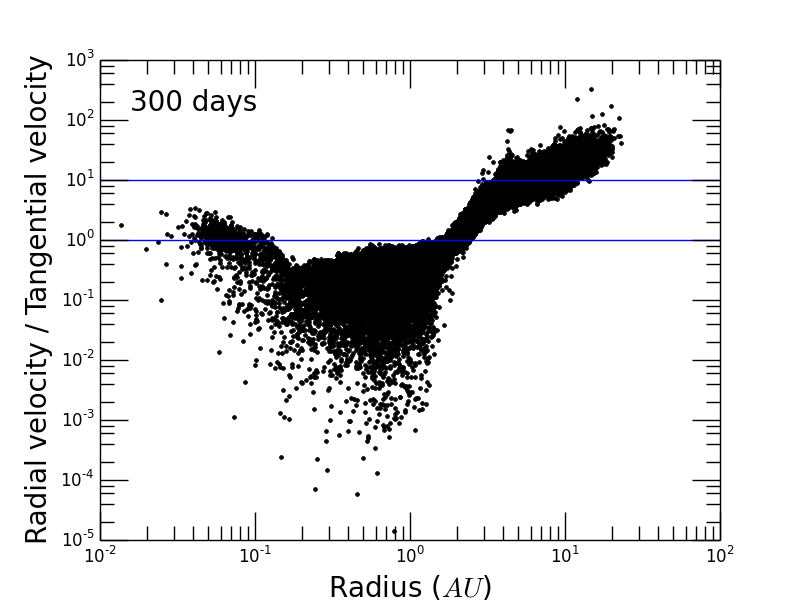}
\includegraphics[scale=0.30, trim=2.5cm 1.0cm 0.0cm 0.0cm]{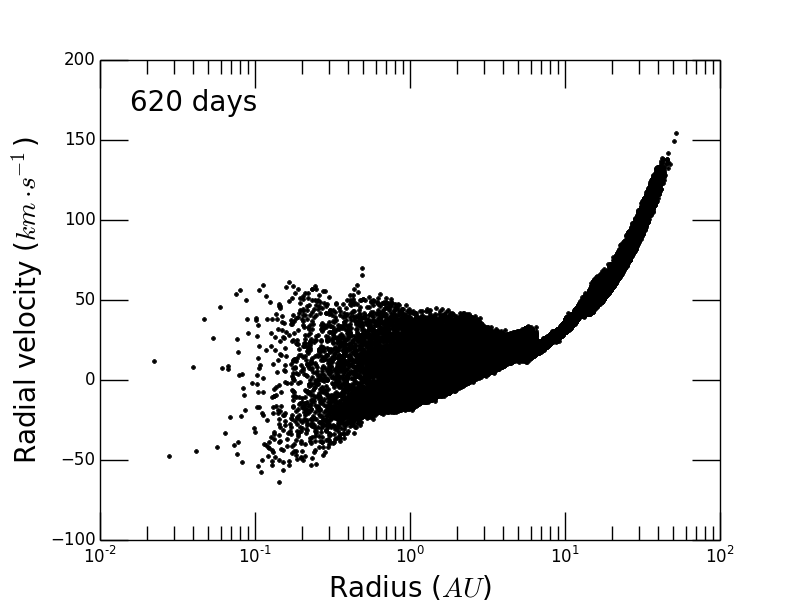}
\includegraphics[scale=0.30, trim=0.0cm 1.0cm 0.0cm 0.0cm]{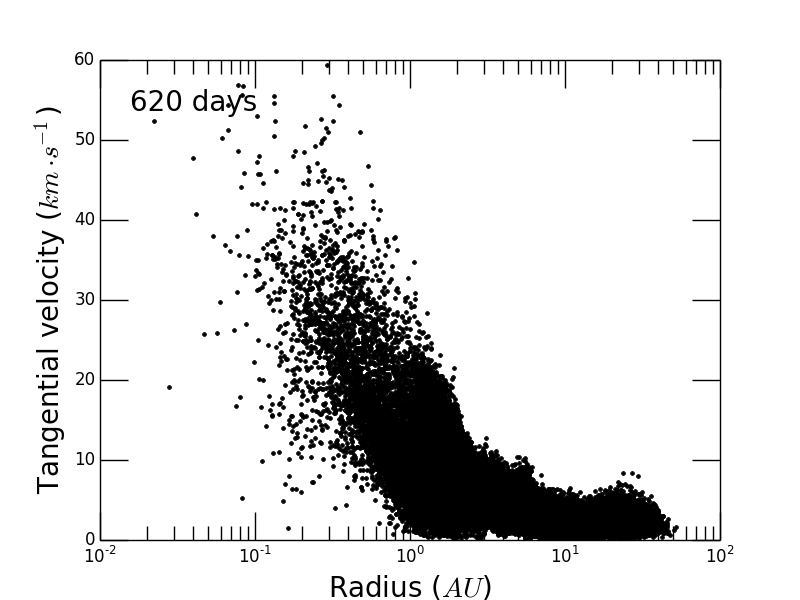}
\includegraphics[scale=0.30, trim=0.0cm 1.0cm 0.0cm 0.0cm]{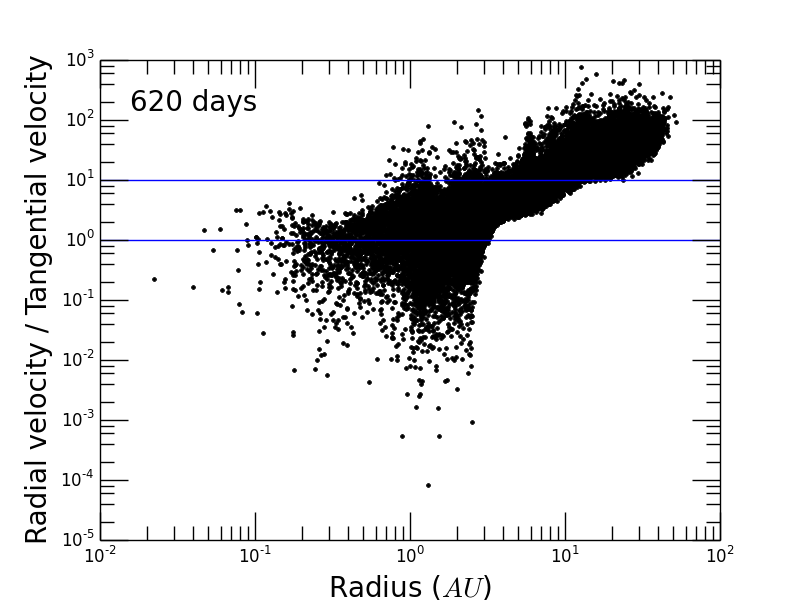}
\includegraphics[scale=0.30, trim=2.5cm 1.0cm 0.0cm 0.0cm]{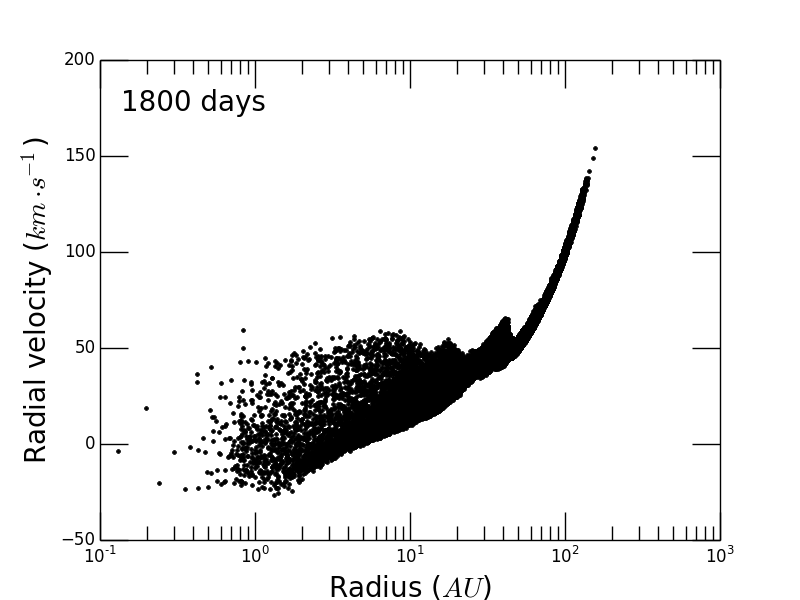}
\includegraphics[scale=0.30, trim=0.0cm 1.0cm 0.0cm 0.0cm]{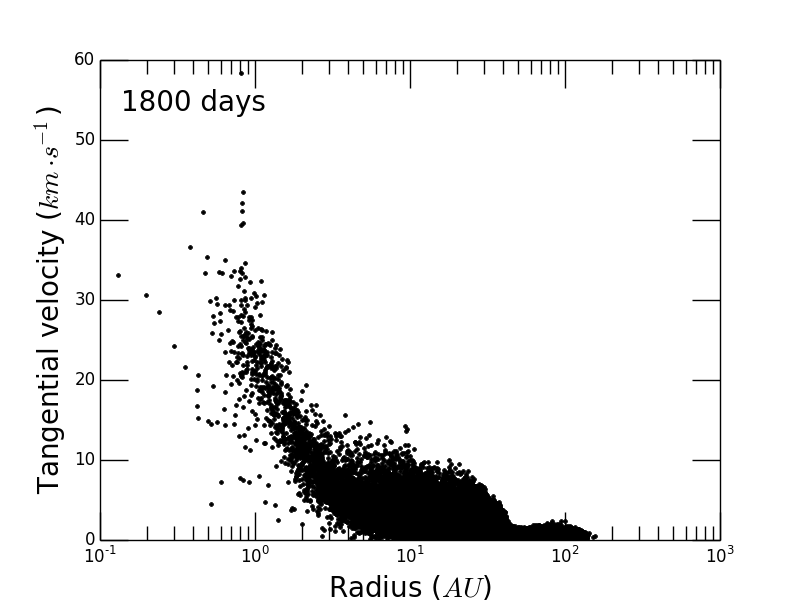}
\includegraphics[scale=0.30, trim=0.0cm 1.0cm 0.0cm 0.0cm]{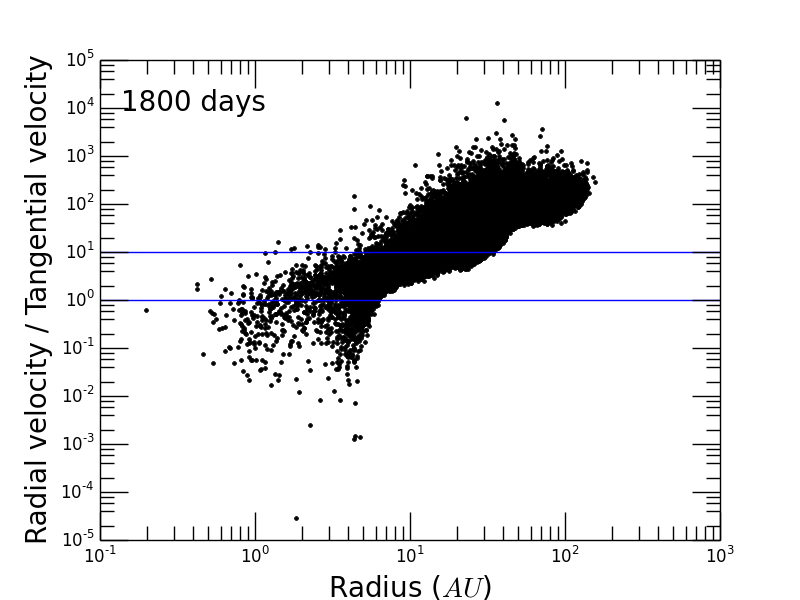}
\includegraphics[scale=0.30, trim=2.5cm 1.0cm 0.0cm 0.0cm]{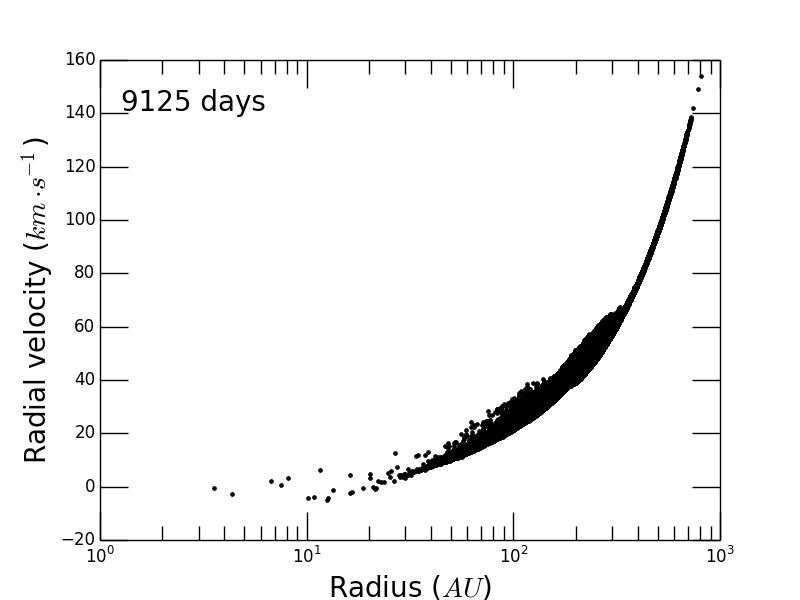}
\includegraphics[scale=0.30, trim=0.0cm 1.0cm 0.0cm 0.0cm]{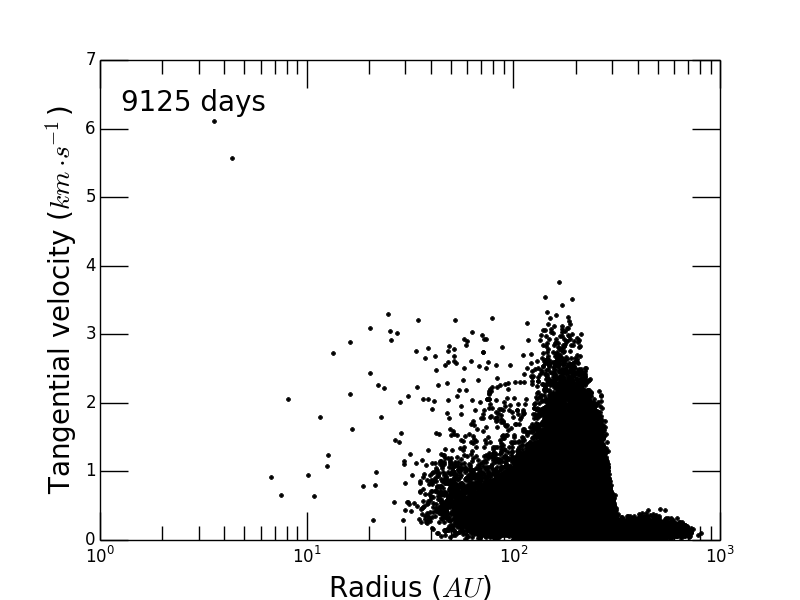}
\includegraphics[scale=0.30, trim=0.0cm 1.0cm 0.0cm 0.0cm]{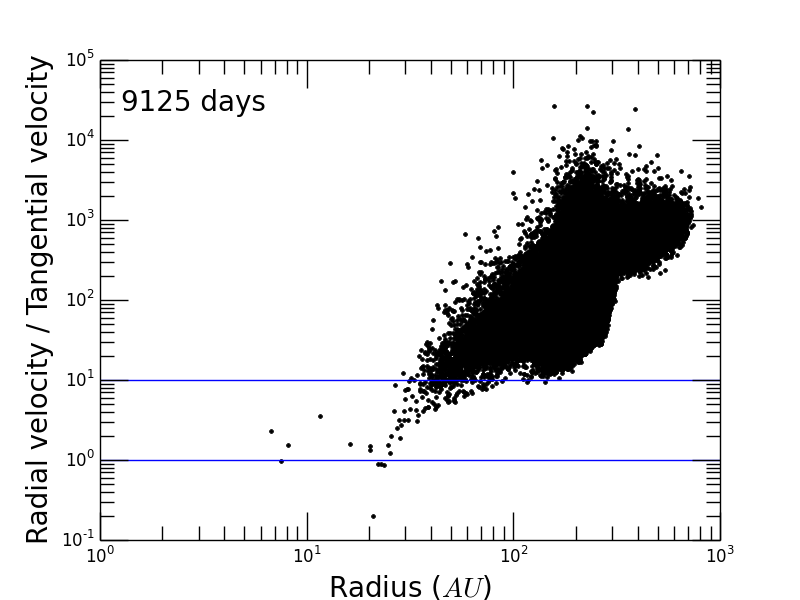}
\includegraphics[scale=0.30, trim=2.5cm 0.0cm 0.0cm 0.0cm]{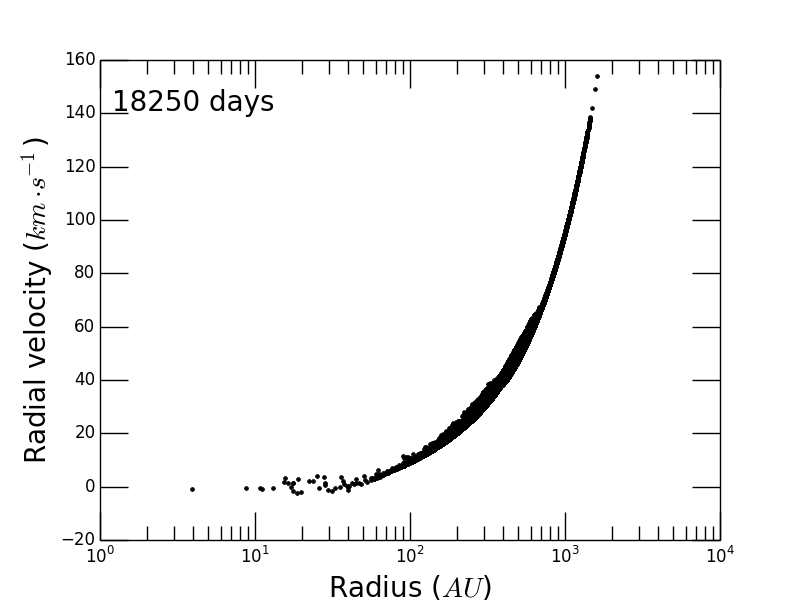}
\includegraphics[scale=0.30, trim=0.0cm 0.0cm 0.0cm 0.0cm]{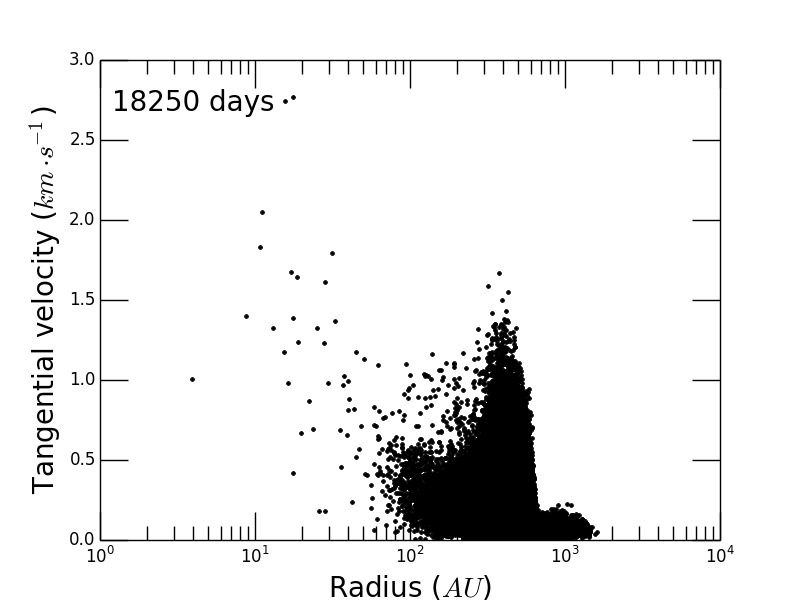}
\includegraphics[scale=0.30, trim=0.0cm 0.0cm 0.0cm 0.0cm]{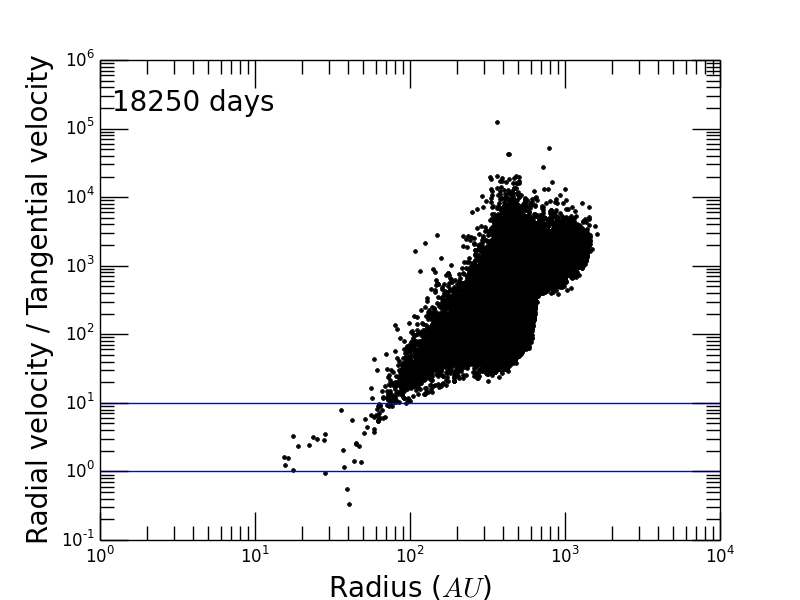}
\caption{\protect\footnotesize{First column: radial velocity vs. radius for the unbound particles. Second column: tangential velocity vs. radius for the unbound particles. Third column: radial over tangential velocity ratio vs. radius for the unbound particles; the horizontal blue lines mark $v_{\mathrm{rad}} > v_{\mathrm{tan}}$ and $v_{\mathrm{rad}} > 10v_{\mathrm{tan}}$, also note that the vertical scale is logarithmic in this column. Rows are in order of increasing time from top to bottom and the snapshots are taken at 300, 620, 1800, 9125 and 18\,250~days, respectively.}}
\label{fig:radial_velocities_multiple_timestep}
\end{figure*}

Homologous expansion assumes gas velocities to be only radial. 
In Figure~\ref{fig:radial_velocities_multiple_timestep}, we compare the radial velocity, $v_{\mathrm{rad}}$, and the tangential velocity $v_{\mathrm{tan}}$ for all  unbound SPH particles at various sample times between 300~days and 18\,250~days to show that this assumption is accurate in our case. The values are reported in Table~\ref{tab:tangential_over_radial}.

\begin{table}
\begin{center}
\begin{tabular}{ccccccc}
\hline
Time   & \multicolumn{3}{c}{$v_{\mathrm{rad}} > v_{\mathrm{tan}}$} & \multicolumn{3}{c}{$v_{\mathrm{rad}} > 10v_{\mathrm{tan}}$}  \\
\cmidrule(r){2-4} \cmidrule(l){5-7} 
(days) & (\#)  & (\ms) & ($M_{\mathrm{env}}$ \%)                   & (\#)  & (\ms) & ($M_{\mathrm{env}}$ \%)                      \\
\hline
300    & 34012 & 0.22 & 45                                         & 8584  & 0.05 & 11                                            \\
620    & 65247 & 0.41 & 86                                         & 22577 & 0.14 & 30                                            \\
1800   & 73971 & 0.47 & 98                                         & 59501 & 0.38 & 78                                            \\
9125   & 75780 & 0.48 & 100                                        & 75645 & 0.48 & 100                                           \\
18250  & 75778 & 0.48 & 100                                        & 75729 & 0.48 & 100                                           \\
\hline
\end{tabular}
\end{center}
 \begin{quote}
  \caption{\protect\footnotesize{Number, mass and mass/$M_{\mathrm{env}}$ of the unbound SPH particles with $v_{\mathrm{rad}} > v_{\mathrm{tan}}$ (2nd, 3rd and 4th columns) and $v_{\mathrm{rad}} > 10v_{\mathrm{tan}}$ (5th, 6th and 7th columns). Here $M_{\mathrm{env}}$ is the envelope mass. The values are given at the sample times used for Figure~\ref{fig:radial_velocities_multiple_timestep}.}} \label{tab:tangential_over_radial}
 \end{quote}
\end{table}

At 300~days, when the orbital separation between the primary's core and the companion has stopped decreasing, only half of the envelope has $v_{\mathrm{rad}} > v_{\mathrm{tan}}$. This fraction quickly increases to $\simeq 86$~percent by 620~days, when almost all recombination energy has been injected in the envelope gas. By 1800 days, $\simeq 98$~percent of the envelope has a radial velocity larger than the tangential component.
A more stringent condition, namely that $v_{\mathrm{rad}} > 10v_{\mathrm{tan}}$, shows that the amount of envelope with a radial velocity at least ten times greater than the tangential velocity is $\simeq 11$~percent at 300~days, when the separation has just stopped decreasing. The amount then becomes $\simeq 78$~percent at 1800~days.
In both cases the SPH particles with the highest tangential velocities reside in the inner envelope, near the central binary formed by the primary's core and the companion, as can be seen in the second and third columns of Figure~\ref{fig:radial_velocities_multiple_timestep}.

We conclude that at 300~days homologous expansion does not reproduce closely a substantial portion of the envelope. By 1800~days about 20 percent of the envelope still does not satisfy our stronger criterion. However, after 1800~days the number of particles satisfying our stronger criterion starts to decrease rapidly and by 5000~days $\simeq 90$~percent of the particles have $v_{\mathrm{rad}} > 10v_{\mathrm{tan}}$. Finally, at 9125~days (equivalent to half of our total simulation time), all SPH particles satisfy the criterion.


\section{Homologous expansion in CE ejecta}
\label{sec:homologous_expansion_in_CE_ejecta}
In this section we apply the homologous expansion kinematic recipe to the unbound ejecta and analyse the similarities and differences between the analytic model and our simulations.

\subsection{Analytic vs. numerical radial velocities}
\label{ssec:analytic_vs_numerical_radial_velocities}
To verify how well the homologous expansion reproduces the kinematics of the ejecta determined by the SPH simulation, we compare the actual SPH particles' radial velocity distribution with that derived from the analytical expression of the homologous expansion model (Equation~\ref{eq:homologous_expansion}), at fixed homologous times.

In supernovae the injection of energy happens over a negligible time compared to the time-scale of the expansion of the ejecta, therefore the initial time, $t_0$, can be set at any time during the energy injection phase, without affecting how the analytical model fits the ejecta kinematics. However, in the case of the CE interaction, the injection of energy happens over a longer time and the choice of $t_0$ must be made more carefully.

\begin{figure*}
\centering     
\includegraphics[scale=0.42, trim=0.8cm 0.0cm 0.8cm 0.0cm, clip]{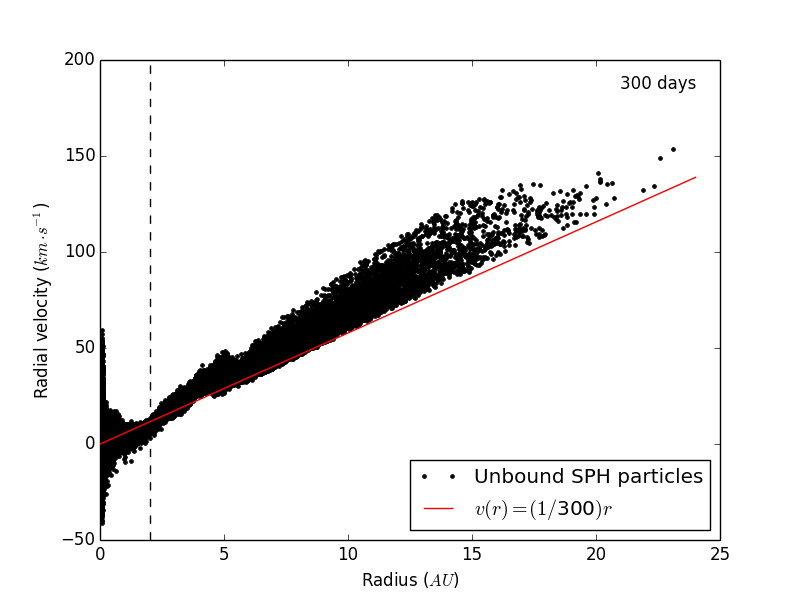}
\includegraphics[scale=0.42, trim=0.8cm 0.0cm 0.8cm 0.0cm, clip]{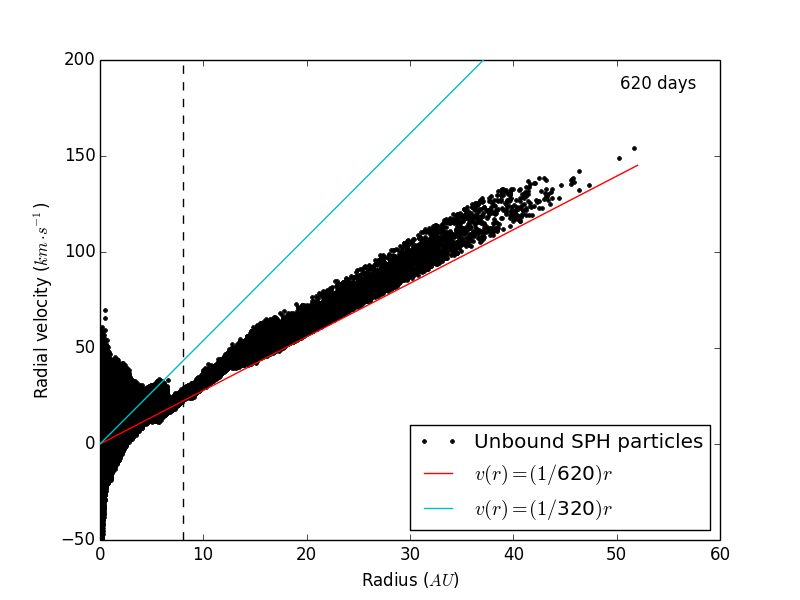}
\includegraphics[scale=0.42, trim=0.8cm 0.0cm 0.8cm 0.0cm, clip]{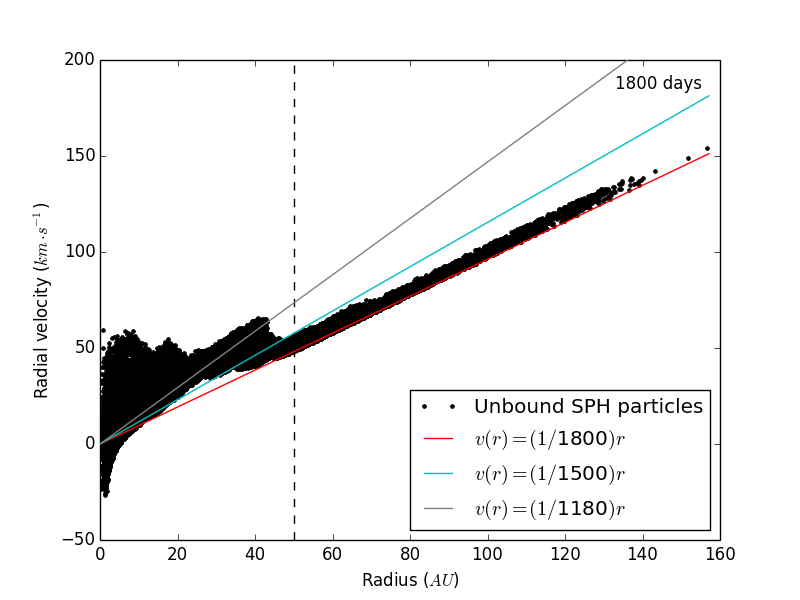}
\includegraphics[scale=0.42, trim=0.8cm 0.0cm 0.8cm 0.0cm, clip]{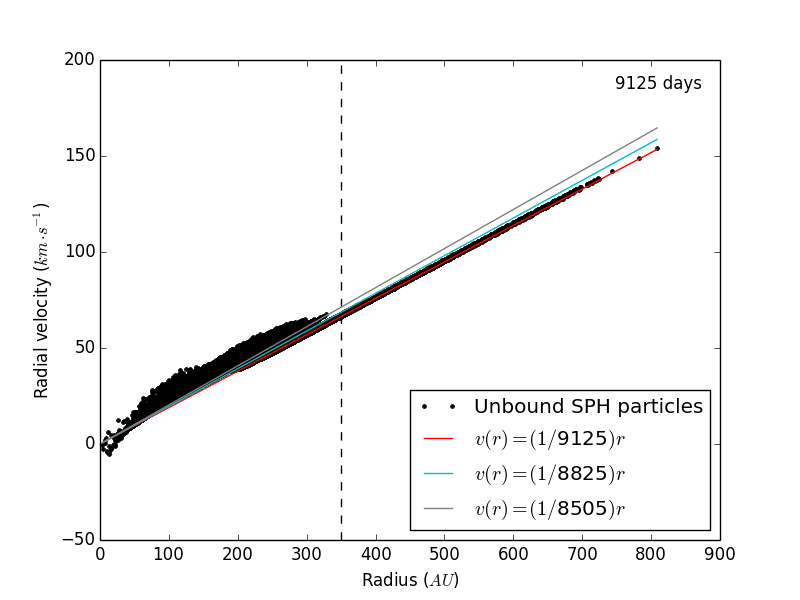}
\includegraphics[scale=0.42, trim=0.8cm 0.0cm 0.8cm 0.0cm, clip]{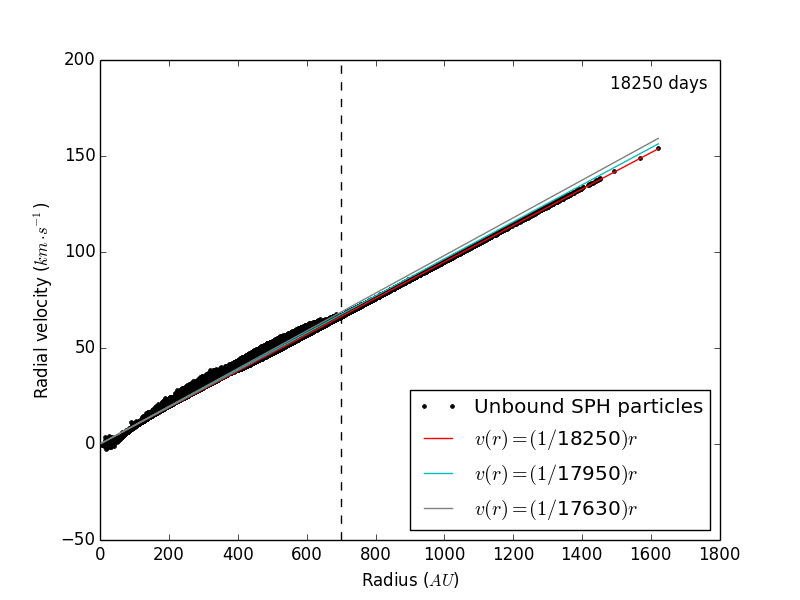}
\caption{\protect\footnotesize{Radial velocity vs. radius for the unbound SPH particles at the same sample times of Figure~\ref{fig:radial_velocities_multiple_timestep} (black dots).
Shown as solid lines are the analytical curves representing the radial velocities predicted by the homologous expansion model, with a range of initial homologous times. The red line corresponds to $t_0 = 0$~days, the cyan line corresponds to $t_0 \simeq 300$~days (i.e., when the orbital energy has been completely injected into the envelope), the grey line corresponds to $t_0 \simeq 620$~days (i.e., when both orbital and recombination energies have been completely injected into the envelope).
Note that the initial homologous time corresponding to the blue model is larger than the time of the first panel, while that corresponding to the grey model is larger than the times of the first two panels.
The vertical dashed lines mark the radius containing the central overdensity of SPH particles described in Section~\ref{ssec:total_mass_and_energy_distributions}.}}
\label{fig:homologous_radius_radial_velocity_fit}
\end{figure*}

We investigate three values for $t_0$: the beginning of the SPH simulation at $t_0 = 0$~days, the end of the rapid in-spiral at $t_0 \simeq 300$~days (i.e., when the orbital energy has been completely injected into the envelope) and the moment when all recombination energy has been injected into the envelope gas at $t_0 \simeq 620$~days. In Figure~\ref{fig:homologous_radius_radial_velocity_fit}, the black dots are the unbound SPH particles, the red, cyan and grey lines are the homologous expansion analytic distributions with $t_0$ = $0$, $\simeq300$ and $\simeq620$~days, respectively.

In the case with $t_0 = 0$~days we observe that as time passes the SPH particles' distribution tends to flatten closer and closer to the analytical distribution, marked as a red line. This is particularly evident for the external portion of the gas distribution (i.e., the part with radius greater than 5~AU in the first panel of Figure~\ref{fig:homologous_radius_radial_velocity_fit}), that gradually moves out in radius and converges with the upper portion of the analytical solution as the simulation advances. The SPH particles closer to the core (i.e., with radius smaller than 5~AU in the first panel of Figure~\ref{fig:homologous_radius_radial_velocity_fit}) also move out in radius and slowly flatten against the analytical line. 

Next let us consider $t_0 = 300$~days and $t_0 = 620$~days. In the second and third panels of  Figure~\ref{fig:homologous_radius_radial_velocity_fit}, we see that the SPH particles do not tend to the analytical distribution. However, the analytical lines pass through the bulk of the particle distribution and possibly could also represent the homologous behaviour at those times.
The asymptotic behaviour of the SPH particles towards the cyan and grey lines in the fourth and fifth panels of Figure~\ref{fig:homologous_radius_radial_velocity_fit} shows that given enough time all the three lines tend to converge. We therefore conclude that, although the energy injection in the case of CE is not instantaneous, on a long enough timescale the choice of the initial time for homologous expansion does not have a great impact on the results. We will therefore adopt $t_0 = 0$~days as the starting point for homologous expansion for the remainder of this paper. As a result of this choice, the homologous time, $t_{\rm h}$, coincides with the elapsed simulation time, $t$. From now on we will use the variable $t$ to identify both simulation and homologous times.

A more detailed study of the correlation between the homologous expansion zero point, $t_0$, and the beginning of the unbinding process would require a set of simulations starting with different configurations tailored not to trigger the dynamic in-spiral as soon as the simulations starts. Examples of such possible configurations are simulations starting at or immediately beyond  Roche lobe overflow (e.g., \citealt{Reichardt2019} and \citealt{Iaconi2017}). However, such configurations require a large amount of computational time, hence we defer this analysis to future work. 

\subsection{Total mass and energy distributions}
\label{ssec:total_mass_and_energy_distributions}
\begin{figure*}
\centering     
\includegraphics[scale=0.42, trim=0.5cm 0.0cm 0.8cm 0.0cm, clip]{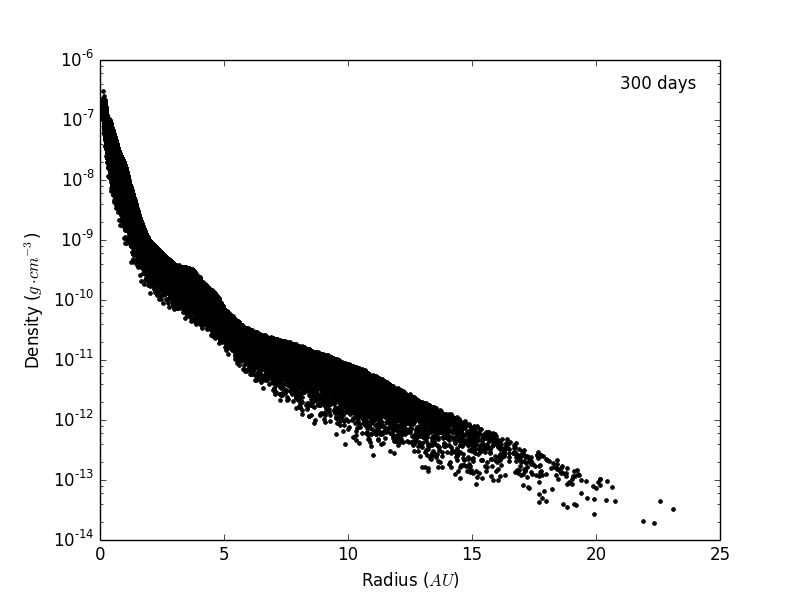}
\includegraphics[scale=0.42, trim=0.5cm 0.0cm 0.8cm 0.0cm, clip]{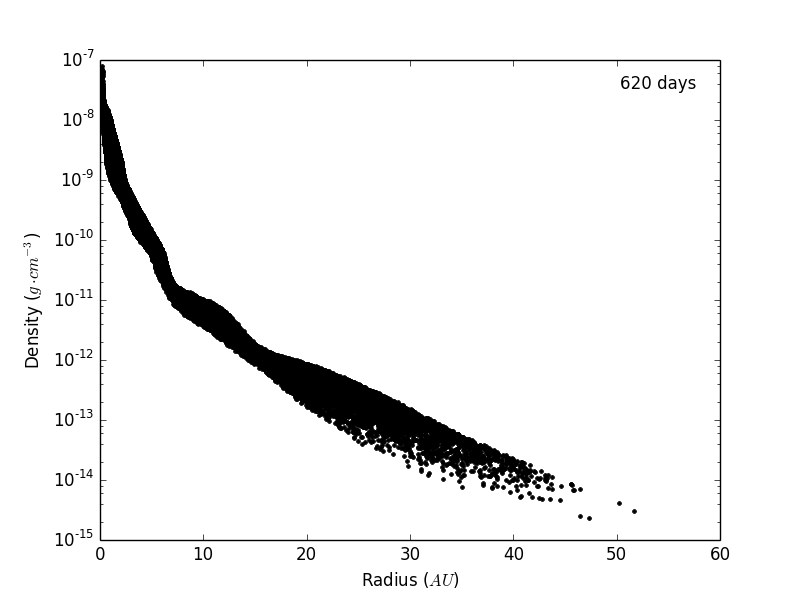}
\includegraphics[scale=0.42, trim=0.5cm 0.0cm 0.8cm 0.0cm, clip]{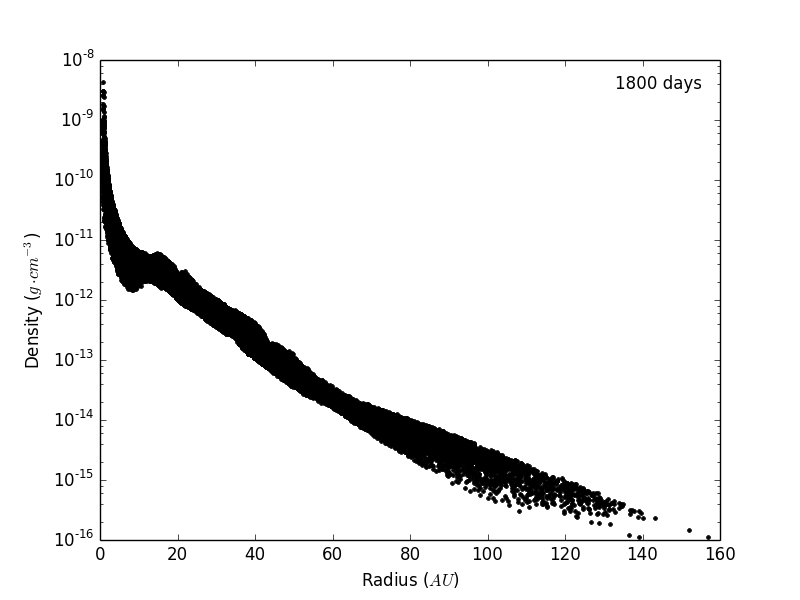}
\includegraphics[scale=0.42, trim=0.5cm 0.0cm 0.8cm 0.0cm, clip]{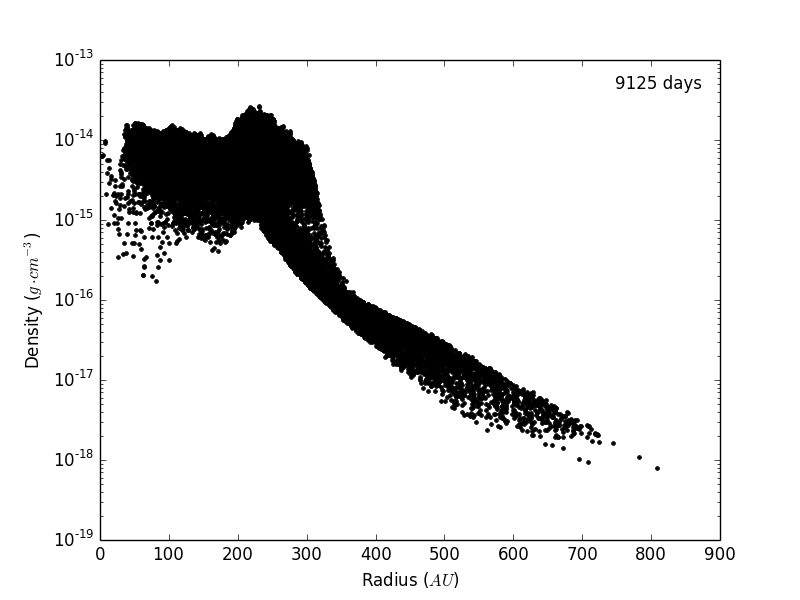}
\includegraphics[scale=0.42, trim=0.5cm 0.0cm 0.8cm 0.0cm, clip]{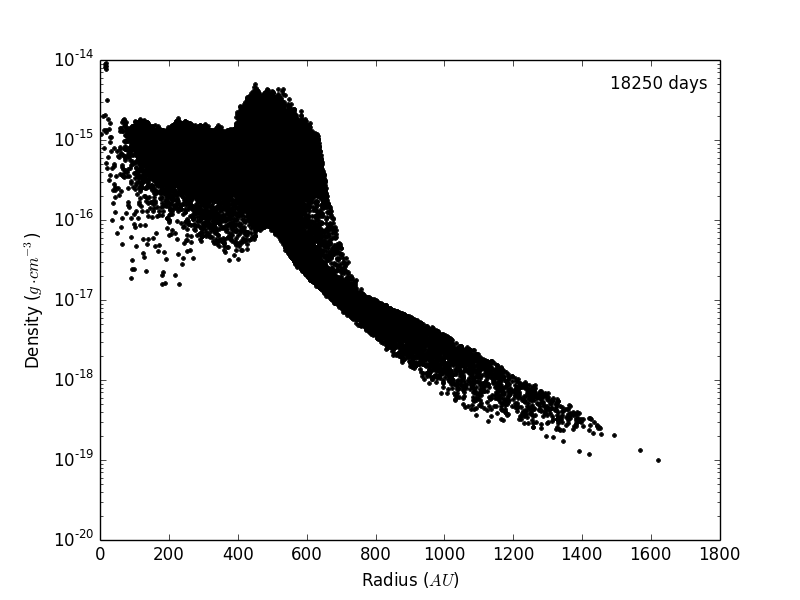}
\caption{\protect\footnotesize{Density vs. radius for the unbound SPH particles at the same sample times of Figure~\ref{fig:radial_velocities_multiple_timestep}.}}
\label{fig:density_vs_radius}
\end{figure*}

In Figure~\ref{fig:homologous_radius_radial_velocity_fit}, first panel, we notice a concentration of SPH particles at radii $\lesssim$2~AU, with radial velocities larger than the homologous model. The concentration can be seen flattening and moving outwards in the remaining panels of Figure~\ref{fig:homologous_radius_radial_velocity_fit} with the outer edge of the concentration moving out to radii $\lesssim$8, $\lesssim$50, $\lesssim$350 and $\lesssim$700~AU for progressive snapshots. We marked this locations with vertical dashed lines in the panels of Figure~\ref{fig:homologous_radius_radial_velocity_fit}. These particles' distribution tends to the homologous distribution as time passes. We find that most of the envelope mass resides in this concentration for the entire duration of the simulation. However, we notice that slowly the external portion of the concentration starts to spread out. This part of the distribution is identifiable as the portion of particles between radii of $\simeq 5$~AU and $\simeq 10$~AU in the second panel of Figure~\ref{fig:homologous_radius_radial_velocity_fit} and $\simeq 25$~AU and $\simeq 60$~AU in the third panel. These particles eventually fully spread out, in terms of mass and energy, mixing with the rest of the envelope at those locations. In Figure~\ref{fig:density_vs_radius} we plot the radial density distributions. The concentration of SPH particles visible in Figure~\ref{fig:homologous_radius_radial_velocity_fit} corresponds to the central density peak.

As the radial velocity plots in Figure~\ref{fig:homologous_radius_radial_velocity_fit} show, initially the SPH particles in the central concentration and those farther out are decoupled, in the sense that the dynamics of the outskirts is already that of a free expanding, homologous system, while the inner particles in the concentration are still interacting with each other and their dynamics is moving towards that of an homologous expanding medium. As time passes the entire central concentration spreads out (note that the range of the abscissa of the panels in Figure~\ref{fig:homologous_radius_radial_velocity_fit} increases with time) and moves towards a more organised radial velocity distribution, even though its density is higher than that the rest of the distribution.

In the last two panels of Figure~\ref{fig:density_vs_radius} we observe that at a fixed radial location within the SPH particle concentration there are SPH particles with a wide range of densities. This is an indication that, as the velocities redistribute, the envelope gas becomes mixed.

We see that this mixing process does not happen on the same time-scale as the dynamical in-spiral and that at $t = 18\,250$~days it is still partially ongoing. Therefore, the process that brings the CE ejecta from an initial velocity distribution, dictated by the complex dynamics of the in-spiral to an ordered distribution such the one described by the homologous expansion approximation, takes place at least over tens of dynamical times.
As introduced in Section~\ref{sec:introduction}, the dynamical in-spiral is powered by a gravitational drag force acting primarily on a local scale approximately the size of the Bondi accretion radius of the companion. This force dissipates energy and angular momentum from the orbit and distributes them to the envelope, also evidenced by the spiral structure imprinted on the ejecta, with gasseous layers expanding away from the central binary primarily on the orbital plane. Therefore the local mechanism that generates a turbulent distribution in both density and velocity, slowly moves towards a symmetric, global distribution. 
In addition, recombination takes place during the first $\simeq620$~days of the simulation and adds an additional expansion force that is reasonably 
symmetric (possibly more with a symmetry axis than spherical symmetry). This force may contribute to circularise the distribution of the envelope and promote the evolution towards a homologous expansion. 

\begin{figure*}
\centering     
\subfigure[]{\includegraphics[scale=0.40, trim=0.0cm 0.0cm 0.0cm 0.0cm]{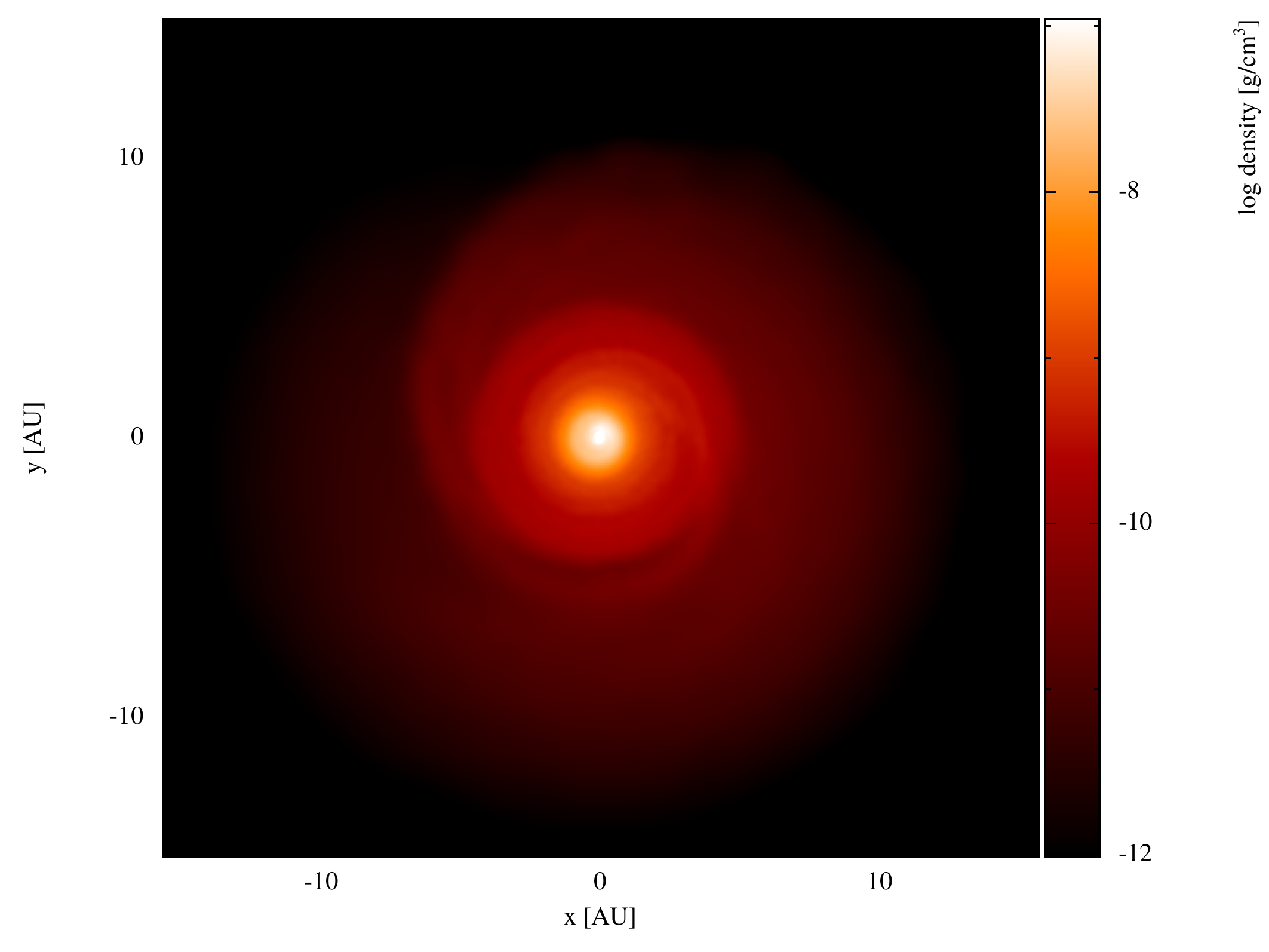}}
\subfigure[]{\includegraphics[scale=0.40, trim=0.0cm 0.0cm 0.0cm 0.0cm]{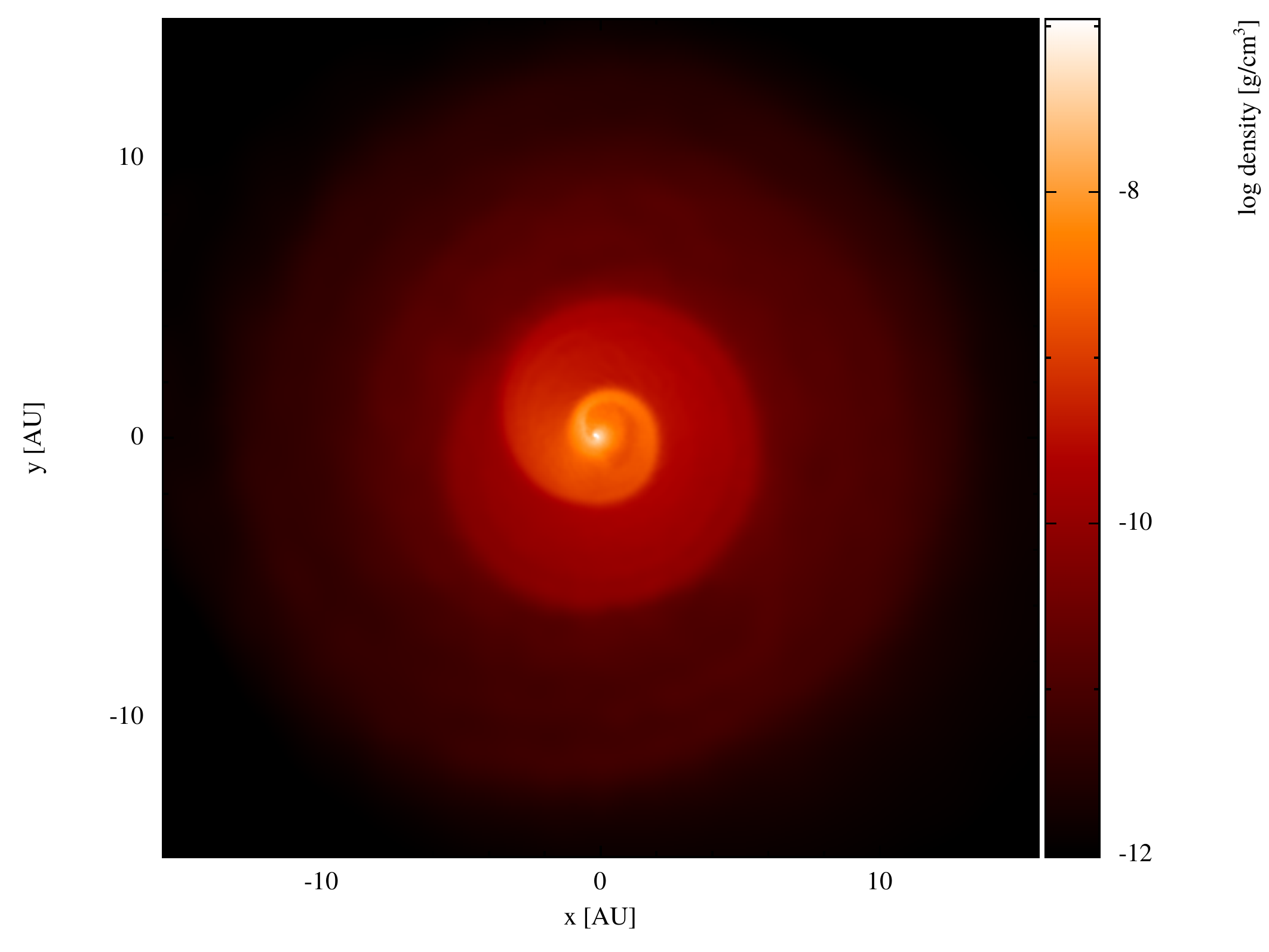}}
\subfigure[]{\includegraphics[scale=0.40, trim=0.0cm 0.0cm 0.0cm 0.0cm]{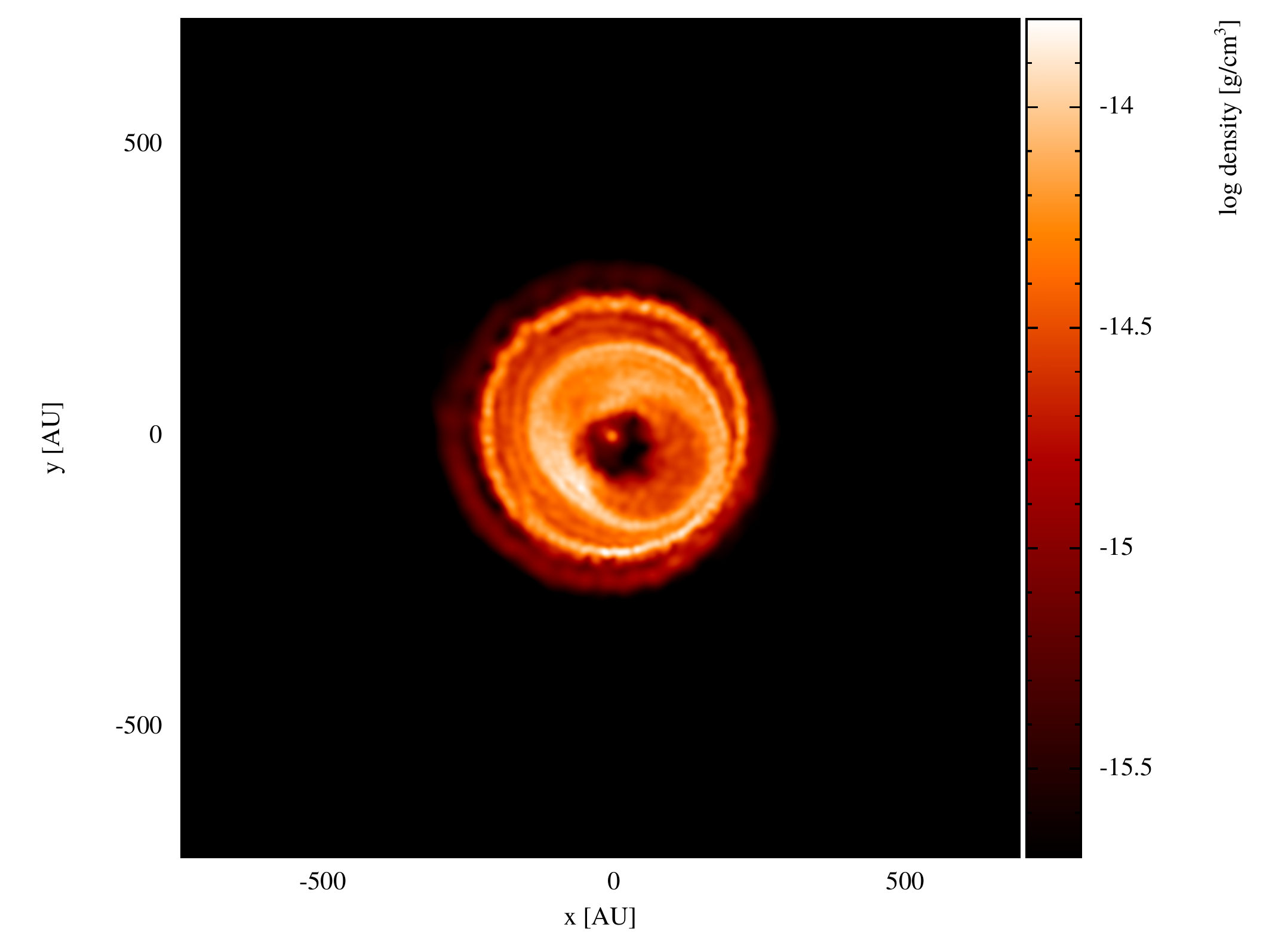}}
\subfigure[]{\includegraphics[scale=0.40, trim=0.0cm 0.0cm 0.0cm 0.0cm]{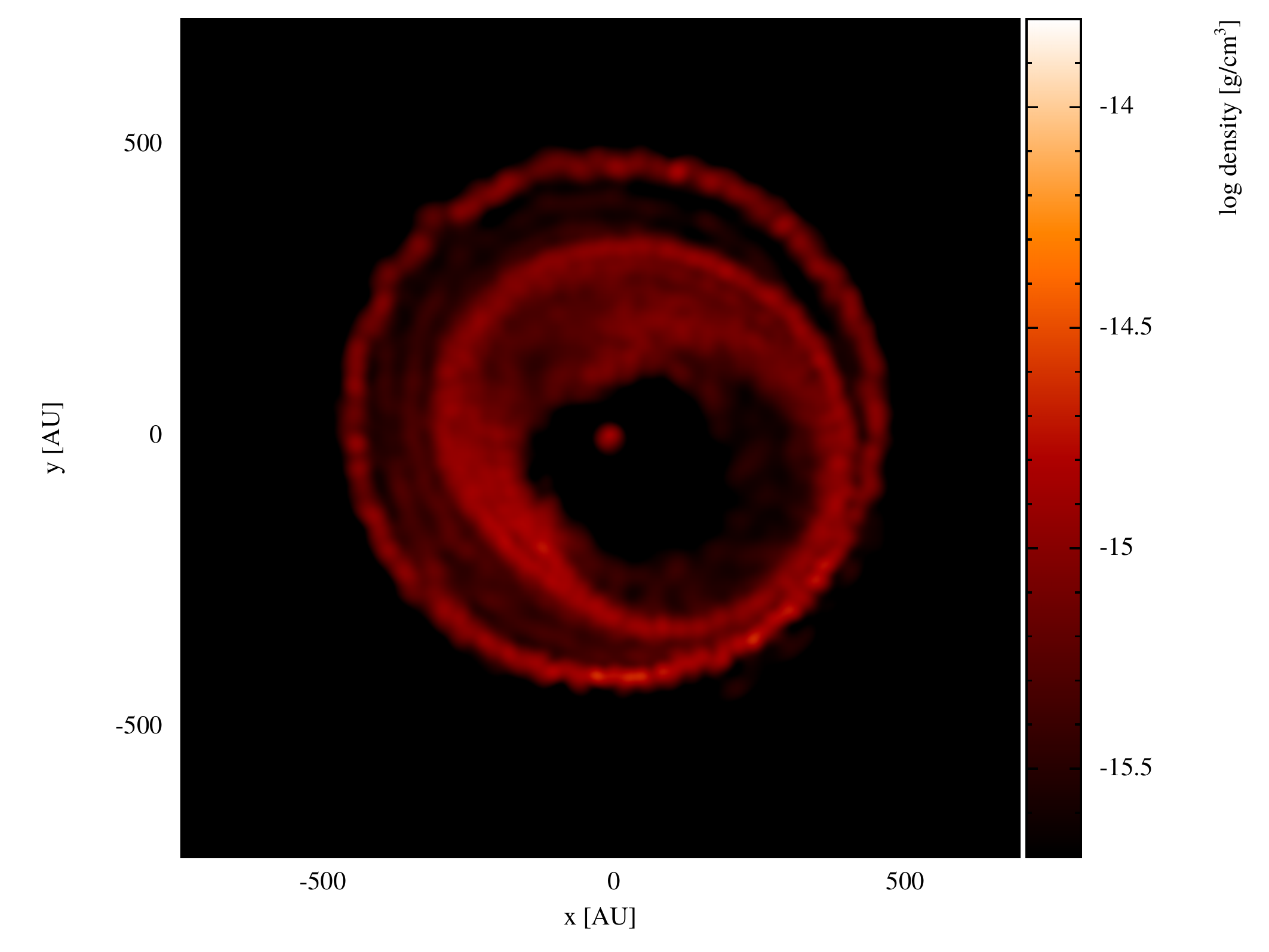}}
\caption{\protect\footnotesize{Density slices along the orbital plane at $z=0$ for the recombination simulation at 300~days (a), 620~days (b), 9125~days (c) and 18\,250~days (d).}}
\label{fig:asymptotic_slices}
\end{figure*}

\begin{figure*}
\centering     
\subfigure[]{\includegraphics[scale=0.40, trim=0.0cm 0.0cm 0.0cm 0.0cm]{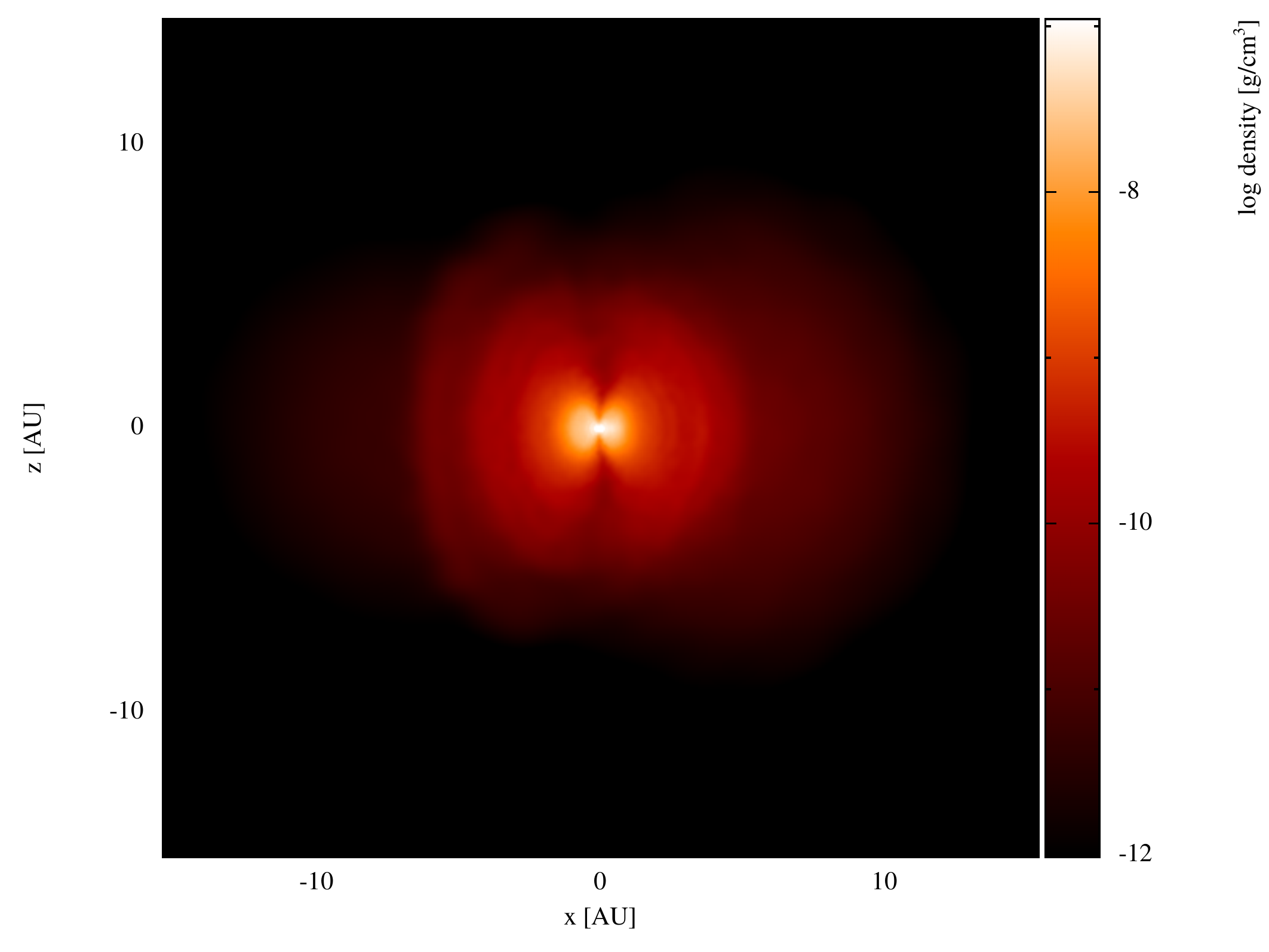}}
\subfigure[]{\includegraphics[scale=0.40, trim=0.0cm 0.0cm 0.0cm 0.0cm]{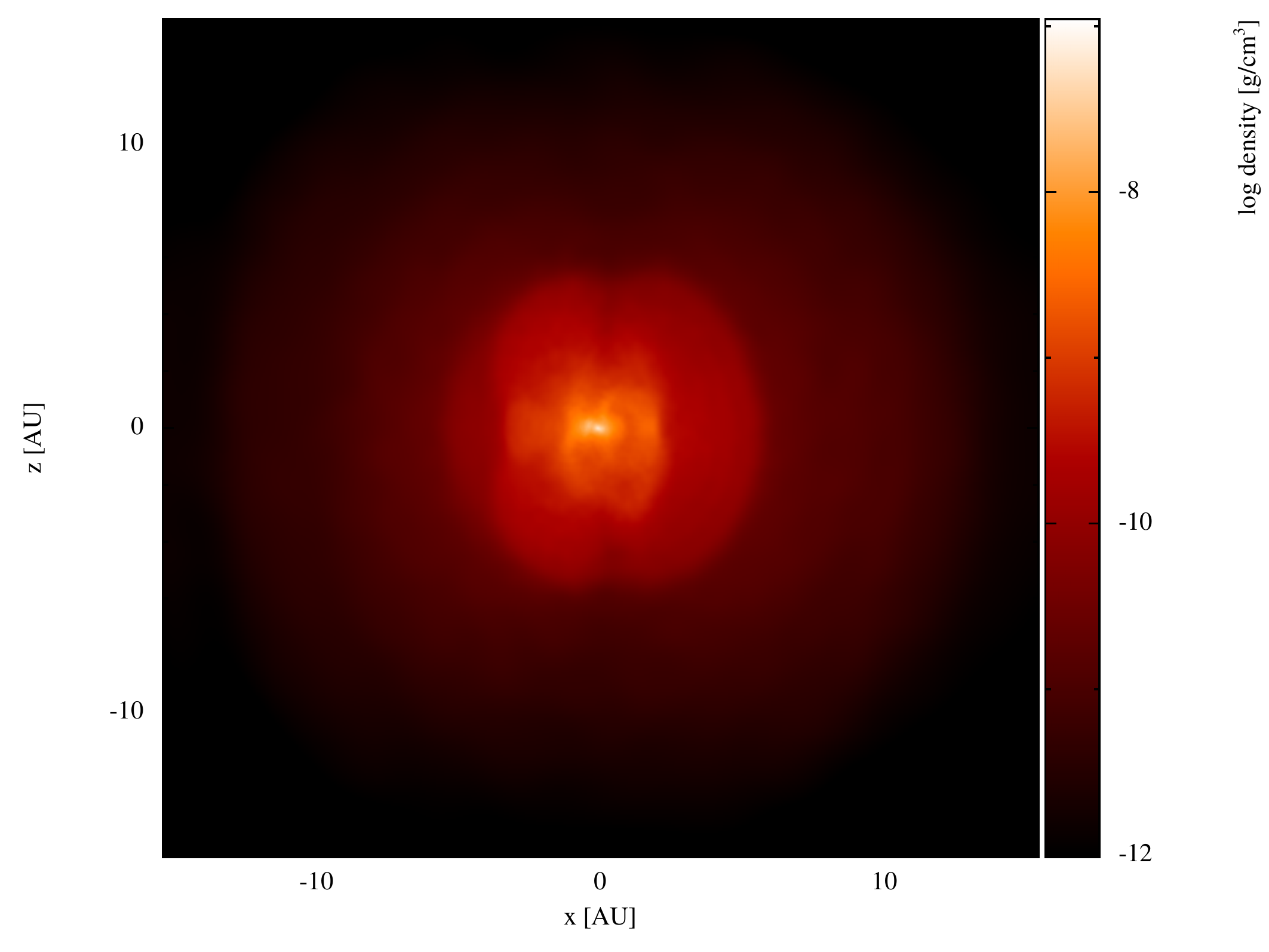}}
\subfigure[]{\includegraphics[scale=0.40, trim=0.0cm 0.0cm 0.0cm 0.0cm]{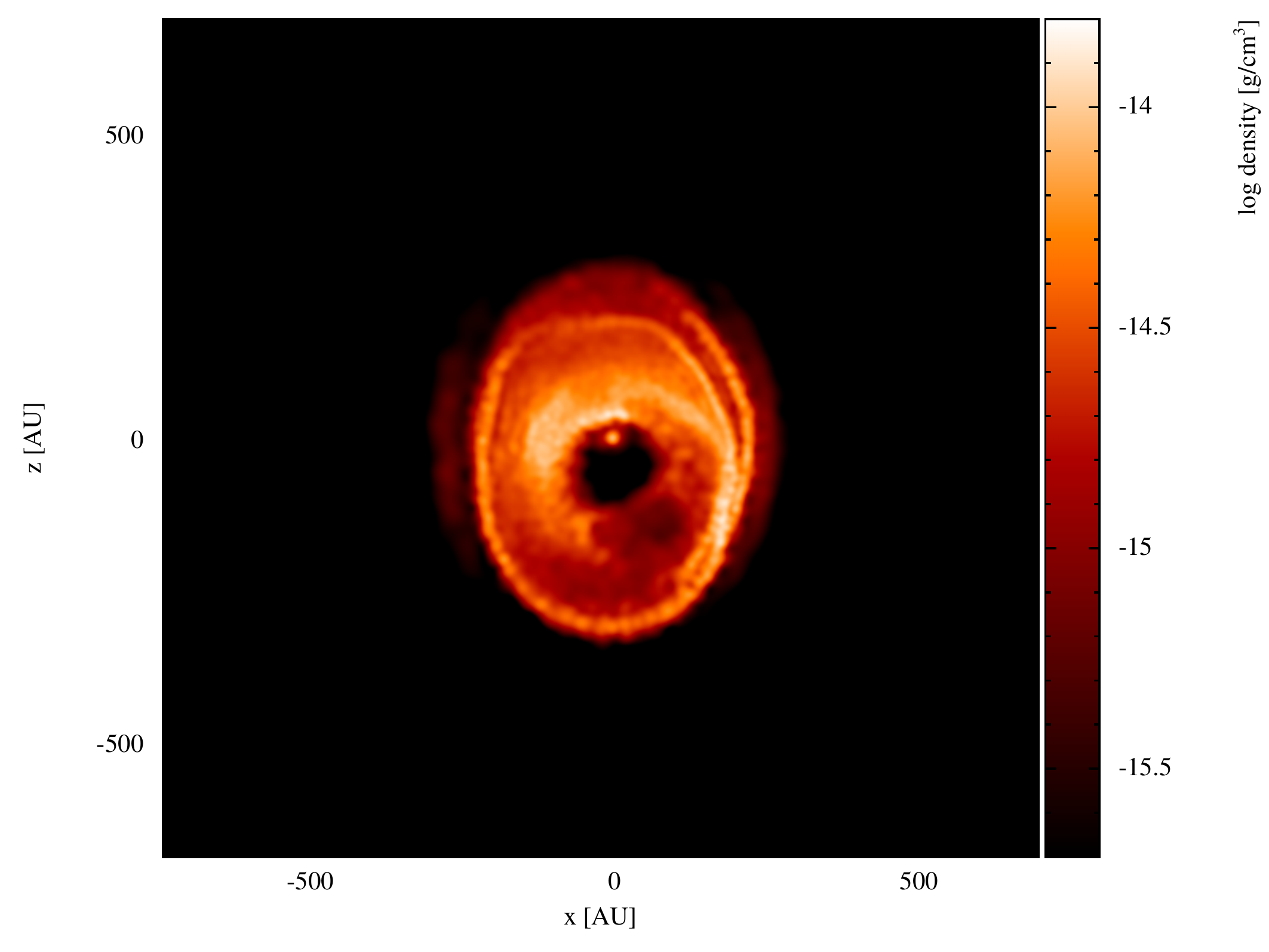}}
\subfigure[]{\includegraphics[scale=0.40, trim=0.0cm 0.0cm 0.0cm 0.0cm]{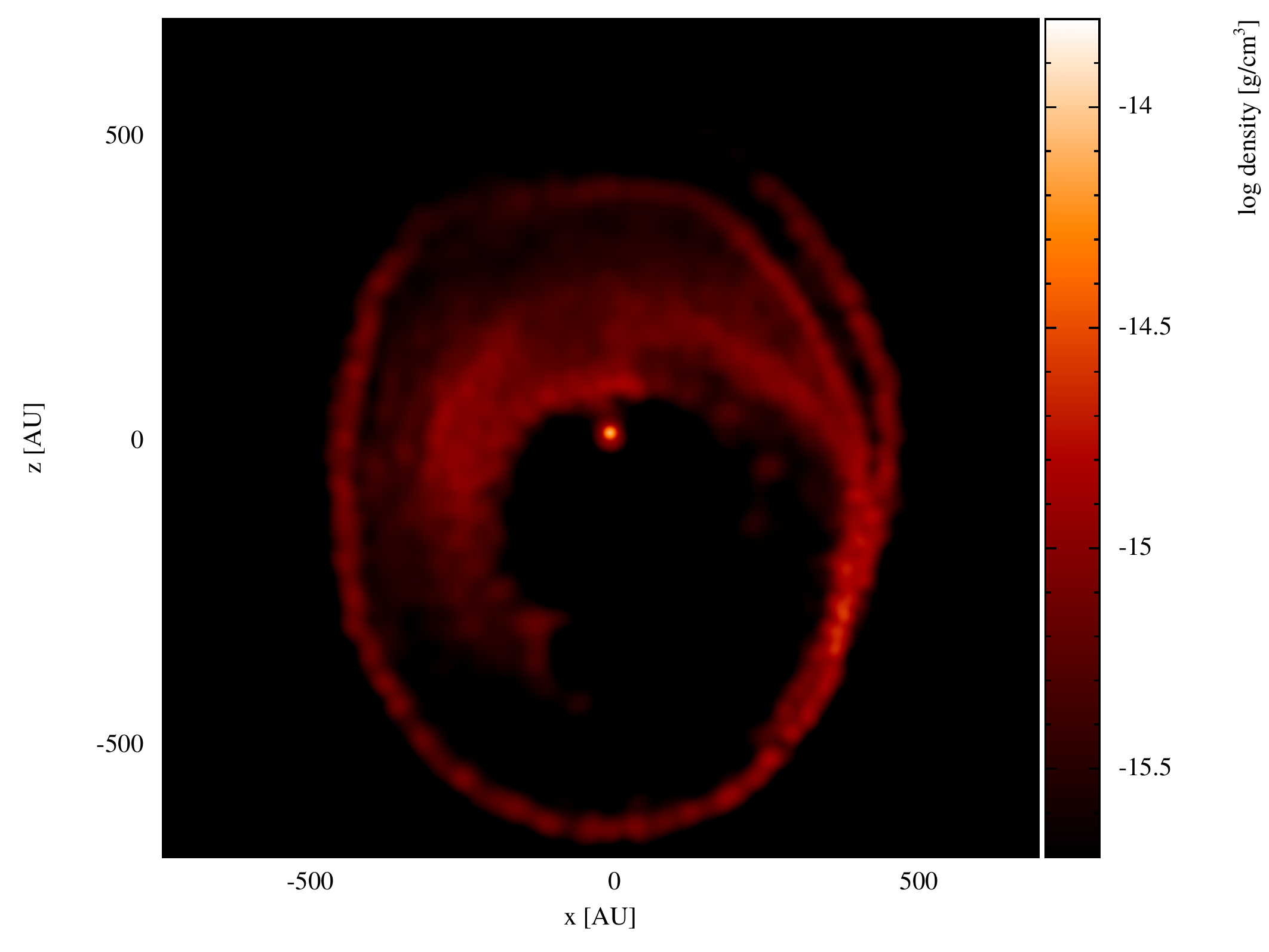}}
\caption{\protect\footnotesize{Density slices along the $x$-$z$ plane at $y=0$ for the recombination simulation at 300~days (a), 620~days (b), 9125~days (c) and 18\,250~days (d).}}
\label{fig:asymptotic_slices_z}
\end{figure*}

\subsection{Shell formation}
\label{ssec:cocoon_formation}
Let us now consider the evolution of the mass and energy contained in the overdensity of SPH particles discussed in Section~\ref{ssec:total_mass_and_energy_distributions}.
By 9125 days we are already in the asymptotic regime where all the energy, both from the orbit and from the gas recombination, has been delivered and the radial velocity distribution is approximately homologous (Figure~\ref{fig:homologous_radius_radial_velocity_fit}).
In the next $9125$~days the central particles' over-density stretches out to $\simeq 350$~au, therefore the velocity at which this particle distribution stretches outwards is $\simeq 66$~km~s$^{-1}$. However, the mass contained within it, i.e., the mass enclosed within 300~au at 9125~days or, 650~au at 18\,250~days, remains approximately the same, $\simeq 0.46$~\ms, and the shape of the distribution itself only shows small changes. An analogous behaviour can be observed for the total energy, which we do not plot.

This is a result of the fact that velocities are almost entirely redistributed at the time of the second-to-last snapshot ($9125$~days), following the homologous expansion kinematics. Hence, we are left with a distribution of material containing a given mass and energy, travelling homologously outwards. In other words, there is an over-density of expanding material shaped like a shell, whose mean density decreases $\propto t^{-3}$ (see Equation~\ref{eq:homologous_density}, and discussion in Section~\ref{ssec:ejecta_cooling}).

Figures~\ref{fig:asymptotic_slices} and \ref{fig:asymptotic_slices_z} show a set of density slices on the orbital plane at $z=0$ and on the $x$-$z$ plane (perpendicular to the orbital plane) at $y=0$, respectively. The slices are taken at 300~days (panel a), 620~days (panel b), 9125~days (panel c) and at 18\,250~days (panel d).
Panels (c) and (d) of Figure~\ref{fig:asymptotic_slices} reveal the disappearance of the spiral pattern typical of the dynamic in-spiral (clear in panels a and b), the cicularisation of the envelope ejecta and the appearance of expanding shell-like features that decrease in density with time. 
Therefore the various outbursts of ejecta generated by each orbit during the dynamic in-spiral tend to separate from one another, as expected from a distribution of gas expanding homologously.
In panels (d) of Figures~\ref{fig:density_vs_radius} and \ref{fig:asymptotic_slices_z} we can also observe that not all the shells manage to perfectly circularise before reaching the homologous regime. This is clear by looking at the inner shell, which maintains an identical shape between 9125~days (panel c) and at 18\,250~days (panel d). Note that, while in panel (c) it is possible to see three shells, the most external, low-density one falls below the density threshold of the figure in panel (d) and is therefore no longer visible.

The overall shape of the expanding shells is elongated in the $z$ direction. This can be explained by the mechanics of the common envelope ejection.
The envelope gas ejected during the dynamic in-spiral is mainly concentrated along the orbital plane (Figure~\ref{fig:asymptotic_slices_z}, panel a), where most of the interaction between sequentially ejected layers of gas happens. The envelope layers ejected at later times impact with the previously ejected ones forming mild shocks (see, e.g., \citealt{Ricker2012}, \citealt{Iaconi2018}). This interaction contributes to slow down the overall expansion velocity on the orbital plane and to diffuse gas along the polar direction. This gas leaves the binary system at different angles outside the orbital plane, allowing the formation of the shell we observe. Additionally, when moving towards the polar direction the gas can expand more freely and maintain a higher velocity, resulting in a shape elongated towards the $z$ axis. 

\citet{Kaminski2018} observed the three red novae V4332 Sgr, V1309 Sco and V838 Mon with ALMA after 22, 8 and 14~yr from their eruptions, time-scales similar to that simulated in this work. The shape of the ejecta of two of them, V4332 Sgr and V1309 Sco, appears bipolar, suggesting therefore that some level of asymmetry such as the one produced in our simulation might be present in systems which are possible post-CE candidates. A more detailed comparison is unfeasible due to many uncertainties and free parameters but it is still interesting to observe such similarity.

The formation of bipolar structures can easily be explained by subsequent phases of ejections where a later ejection impacts the equatorially enhanced CE outflow. This is well known to give rise to bipolar planetary nebulae around post-CE binaries \citep{Frank2018}.

\subsection{Adiabatic cooling of the ejecta}
\label{ssec:ejecta_cooling}
\begin{figure*}
\centering
\subfigure[]{\includegraphics[scale=0.40, trim=0.0cm 0.0cm 0.0cm 0.0cm]{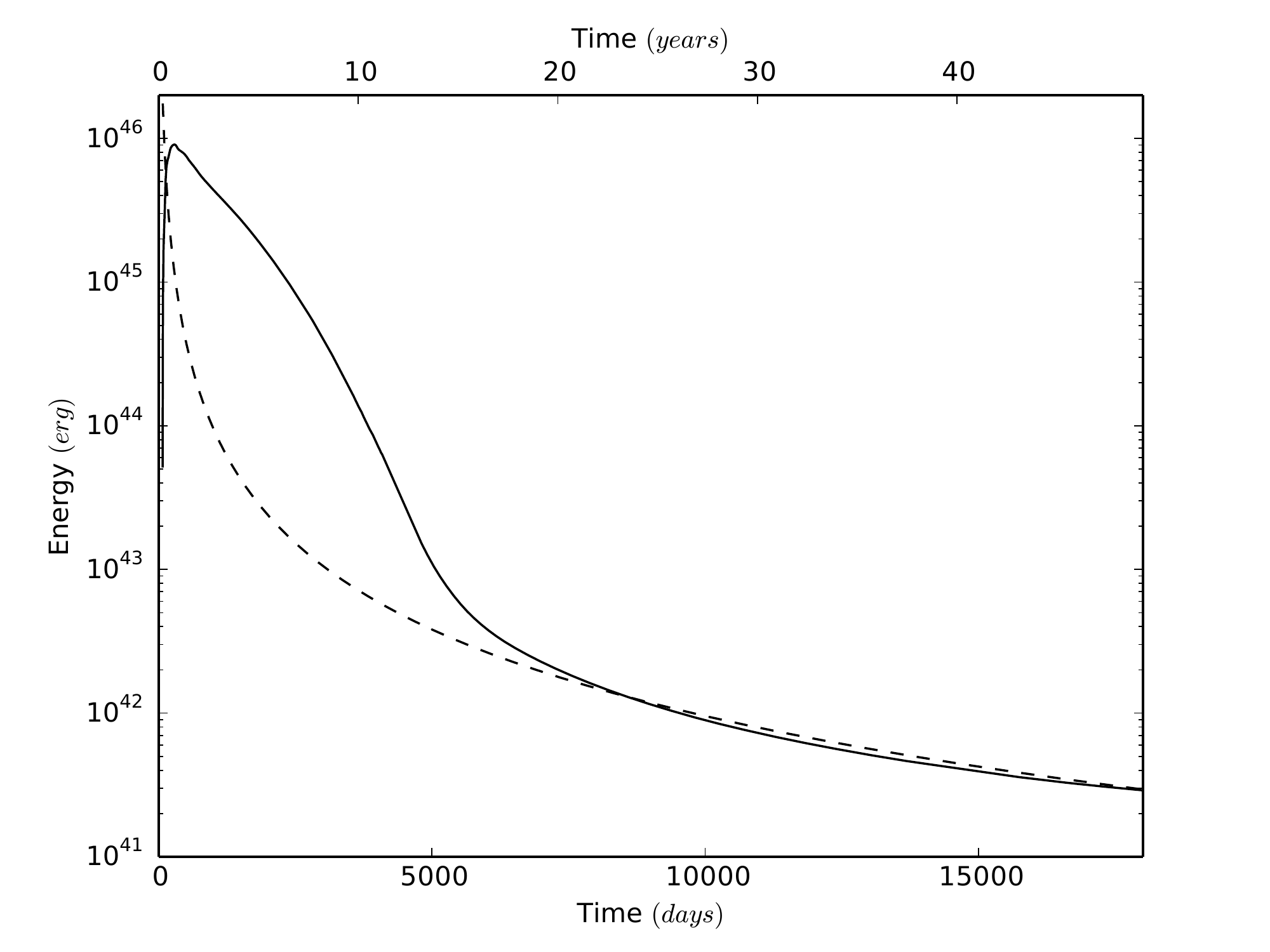}}
\subfigure[]{\includegraphics[scale=0.40, trim=0.0cm 0.0cm 0.0cm 0.0cm]{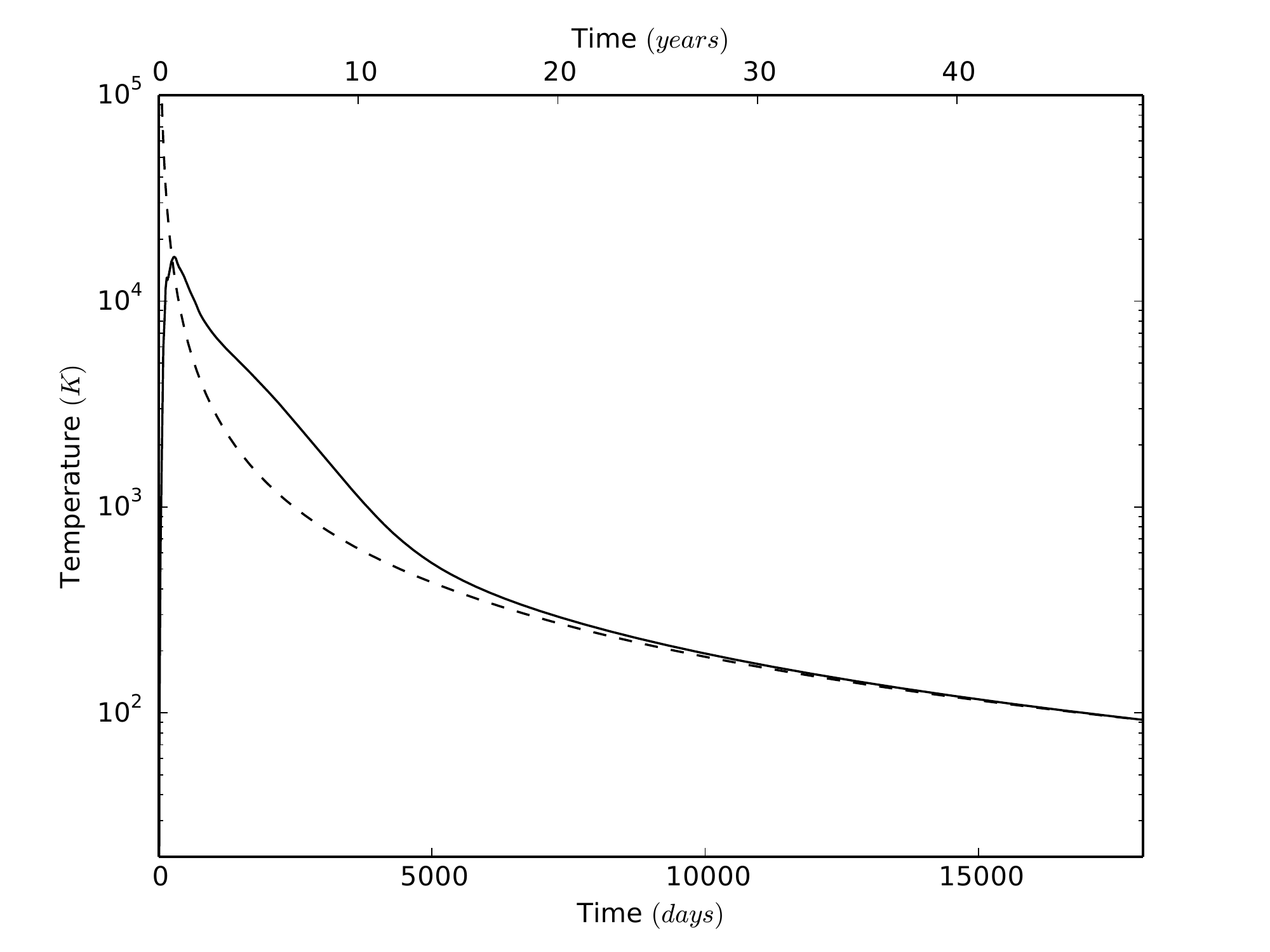}}
\subfigure[]{\includegraphics[scale=0.40, trim=0.0cm 0.0cm 0.0cm 0.0cm]{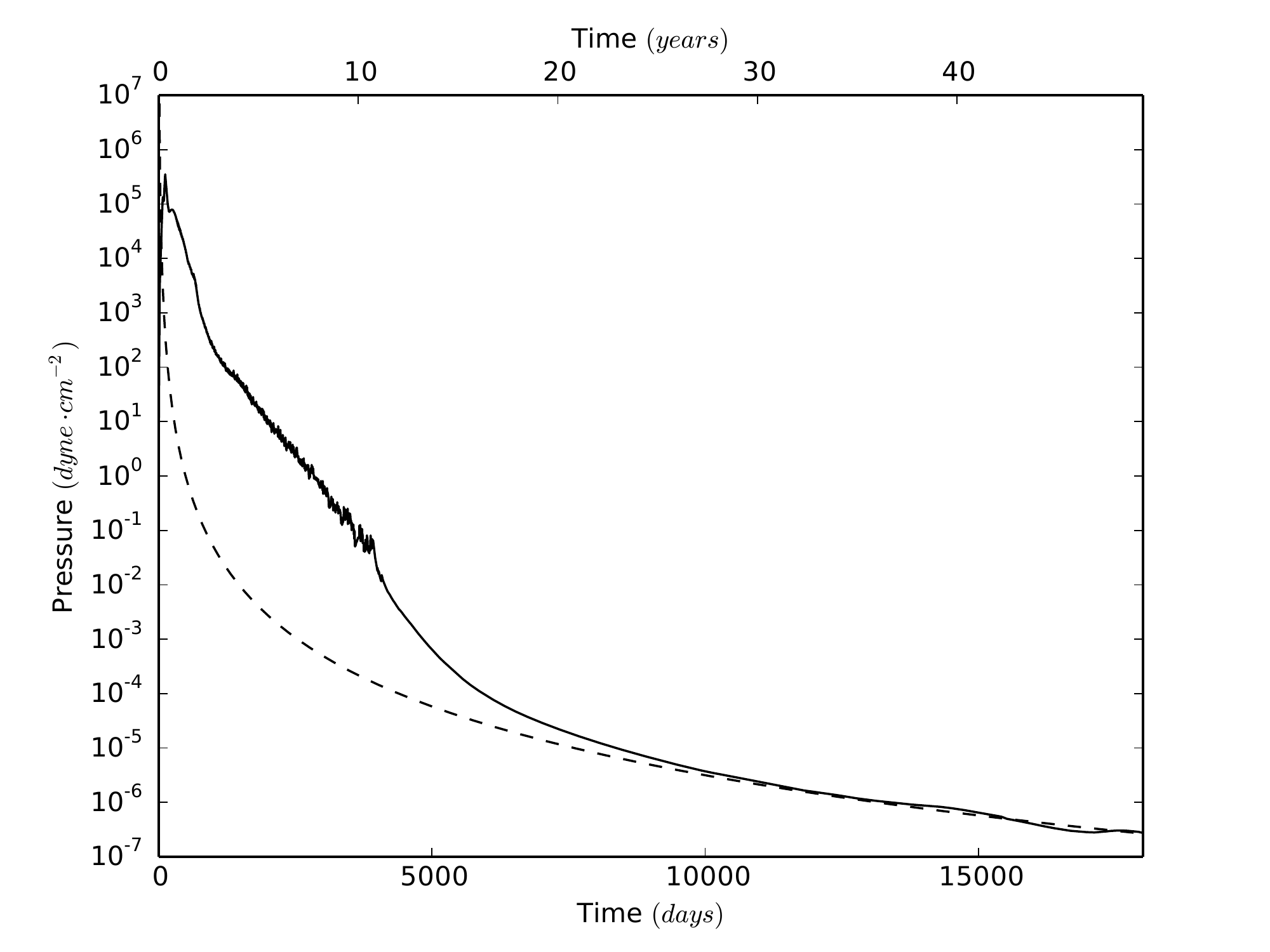}}
\subfigure[]{\includegraphics[scale=0.40, trim=0.0cm 0.0cm 0.0cm 0.0cm]{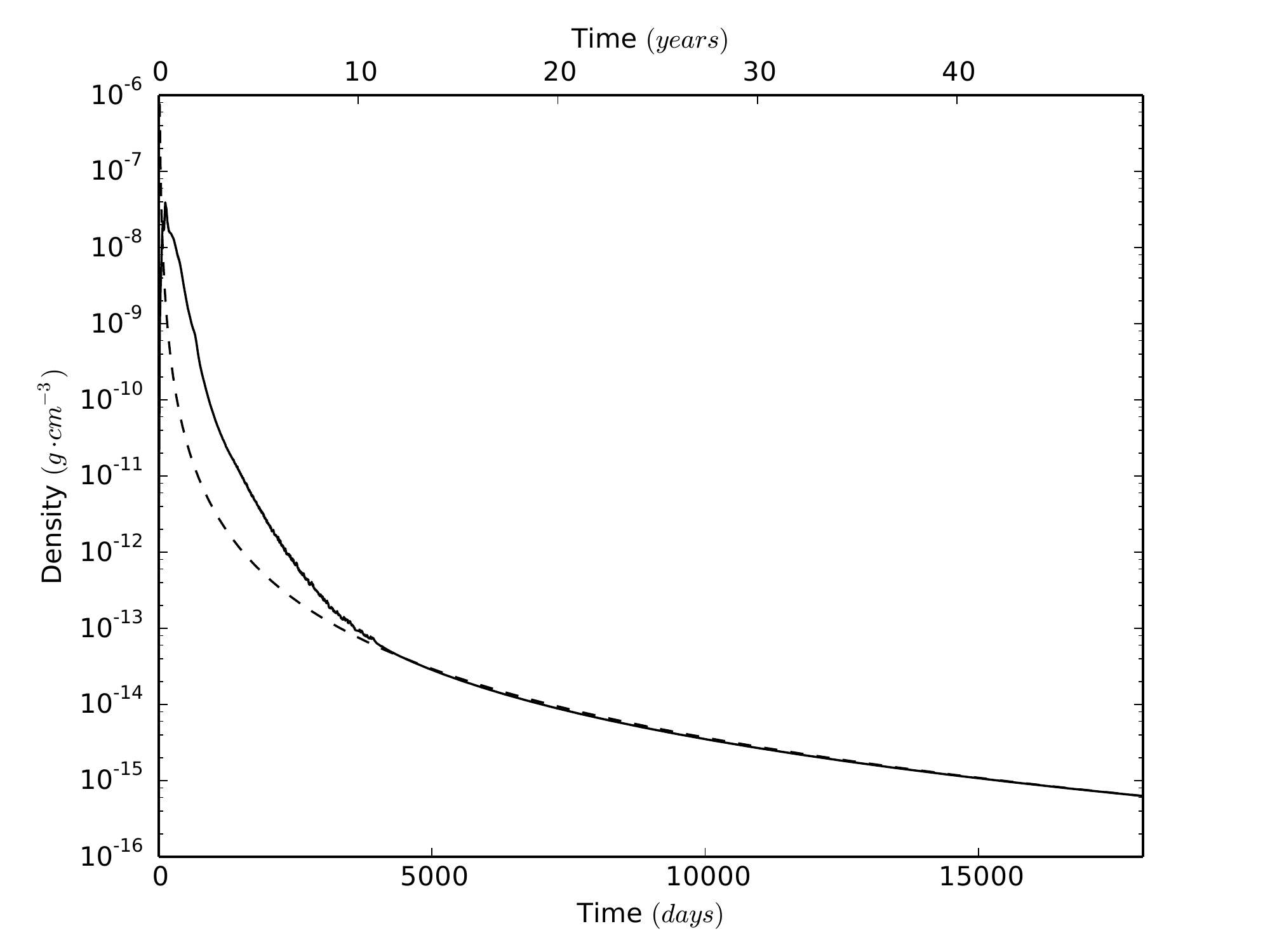}}
\subfigure[]{\includegraphics[scale=0.40, trim=0.0cm 0.0cm 0.0cm 0.0cm]{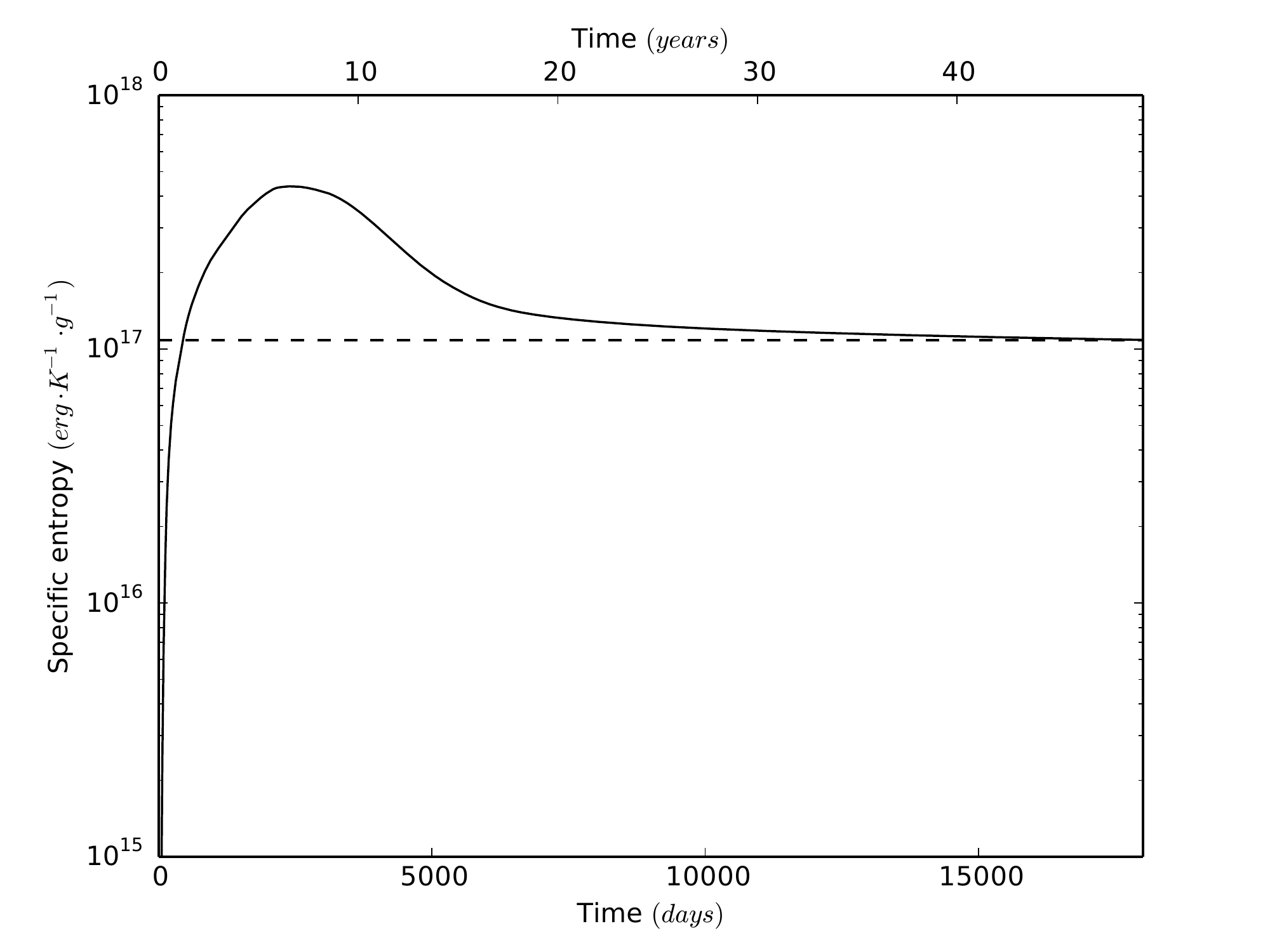}}
\caption{\protect\footnotesize{Thermal energy (a), average temperature (b), average pressure (c), average density (d) and average entropy (e) vs. time for the unbound particles. 
Additionally, we report as dashed lines the curves representing the temporal evolution of the quantities according to the analytic derivation of Section~\ref{sec:numerical_choices_and_model_sanity_checks}. We determined the proportionality constants of the analytic curves so as to match the numerical values at the end of the simulations. Namely $t_{\mathrm{f}} = 18250$~days, $E_{\textrm{therm,f}} \simeq 2.8 \times 10^{41}$~erg, $T_{\mathrm{f}} \simeq 89.7$~K, $P_{\mathrm{f}} \simeq 2.4 \times 10^{-7}$~dyne~$\cdot$~cm$^{-2}$, $\rho_{\mathrm{f}} \simeq 5.8 \times 10^{-16}$~g~$\cdot$~cm$^{-3}$ and $S_{\mathrm{f}} \simeq 1.1 \times 10^{17}$~erg~$\cdot$~K$^{-1}$~g$^{-1}$, where the subscript \virg f'' denotes the values from the last output of the simulations.}}
\label{fig:various_vs_time_100rs_phantom_mesa}
\end{figure*}

Let us now examine how the adiabatic condition is satisfied when the ejecta cool down as they approach the homologous expansion regime.
The results are shown in Figure~\ref{fig:various_vs_time_100rs_phantom_mesa}, where we plot thermal energy (panel a), average temperature (panel b), average pressure (panel c), average density (panel d) and specific entropy (panel e) for the unbound ejecta. All the averages are arithmetical, i.e., we summed the quantities over all the particles then divided by the particle number (this has the effect of mass-weighing the averages because SPH particles have equal mass).
We overplotted (dashed curves) the analytical behaviour of the various physical quantities according to the adiabatic expansion law under homologous conditions, as shown in Equations~\ref{eq:homologous_thermal_energy} to \ref{eq:homologous_entropy}. 

The total thermal energy of the unbound ejecta (Figure~\ref{fig:various_vs_time_100rs_phantom_mesa}, panel a) increases more and more as mass becomes progressively unbound in the first $\simeq 620$~days. This increase is followed by a decrease, steeper at first and then asymptotically flattening. We find a very good agreement between the computed energy and the analytic expression (dashed black line) after the end of the steep decrease at $\simeq 5000$~days. This indicates that the expansion becomes adiabatic after the initial part of the interaction.

The average temperature shows a behaviour in line with the energies and a good agreement with the analytic expectation, in accordance with the fact that temperature and energy are directly proportional for an ideal polytropic gas (Equation~\ref{eq:ideal_gas_etherm}). Finally, also pressure and density have a similar pattern as those of the previous thermodynamic variables, since their analytic evolution can be derived from the equation of state once temperature is known. The numerical values tend to converge with the dashed lines after $\simeq 5000$~days.

A special mention goes to the entropy evolution. Entropy depends on the behaviour of all the other thermodynamic quantities discussed above, in fact, in our model it can be derived from the fundamental thermodynamic relation
\begin{equation}
 dE_{\mathrm{therm}} = TdS - PdV\ ,
 \label{eq:fundamental_thermodynamic_relation}
\end{equation}
where $V \propto M/\rho$. Therefore entropy is very sensitive to deviations from an adiabatic behaviour. We observe that also entropy converges towards the analytic solution after $\simeq 5000$~days, confirming the previous results. We are hence witnessing an expansion that tends to the behaviour of an ideal, polytropic gas under adiabatic conditions.

Our simulation, started with the companion on the stellar surface, ignores the possible effects that circumstellar material ejected prior to the dynamic in-spiral might have on the entropy generation. The envelope lost through the Lagrangian point L2 before the dynamic in-spiral (\citealt{Pejcha2017}, \citealt{Reichardt2019}) may be shocked by the higher velocity ejecta generated by the dynamic in-spiral and result in an additional entropy increase with respect to the one observed in Figure~\ref{fig:various_vs_time_100rs_phantom_mesa} (panel e). Such entropy increase would in any case happen on time-scales of at least one order of magnitude smaller than those required to reach homologous expansion and would not affect the asymptotic evolution of the system.

The entropy evolution is also in line with what was observed by \citet{Ivanova2016} and \citet{Ivanova2018}, with an initial entropy increase at the hand of the drag force energy deposition and shocks, followed by a decrease of the gas entropy while recombination takes place during the expanding phase after the dynamic in-spiral. Eventually, the gas enters the asymptotic regime, where entropy becomes constant, which takes place when the already recombined layers of expanding gas gradually stop interacting with each other and tend towards homologous expansion. 



\section{Characterisation of the photosphere}
\label{sec:determination_of_the_photosphere}
In this section we provide a simple estimate of the location of the photosphere along the orbital plane. The calculation we perform below could be carried out on the computational data without assuming that the envelope expands following the homologous expansion law. However, here we couple the procedure for  determining the photosphere's location with homologous expansion, which, as we have shown in Sections~\ref{sec:numerical_choices_and_model_sanity_checks} and \ref{sec:homologous_expansion_in_CE_ejecta}, approximates the ejecta evolution at large times after the dynamic in-spiral has completed. This is to show how using the analytical homologous pattern to model the ejecta evolution opens the possibility of 
locating and characterising the photosphere and the ejecta emission properties for much longer times than can be calculated in 3D.

A more detailed determination of the photosphere location would require a dedicated radiation transfer simulation assuming the same homologous pattern. We defer such study to future work.

\subsection{Method}
\label{ssec:method}
Assuming now that after $\simeq 5000$~days the envelope kinematics follows the homologous expansion, we can utilise the physical quantities from the code at 5000~days, coupled with the analytical recipe, to create a simple model to estimate location and temperature of the photosphere as a function of time. We assume that the light emitted by the central close binary system cannot be seen directly as it is well within the ejecta, so that the light is reprocessed by the ejecta and emitted as a black-body at the photosphere. Under such conditions we can apply the Stefan-Boltzmann law to our photosphere to determine its temperature:
\begin{equation}
T_{\mathrm{ph}} = \Big( \frac{L}{4 \pi R_{\mathrm{ph}}^2 \sigma} \Big)^{\frac{1}{4}} \ ,
 \label{eq:stefan_boltzmann_law}
\end{equation}
where $L$ is the central object's luminosity, $R_{\mathrm{ph}}$ is the radius of the photosphere and $\sigma$ is the Stefan-Boltzmann constant. 

For the luminosity of the central object we assumed a minimum of $L_{\mathrm{core}}$ and a maximum of $L_{\mathrm{Edd}}$. $L_{\mathrm{core}}$ is the luminosity of the RGB core at the moment when the CE terminates. The luminosity of the companion can be neglected if the object is a main sequence star or a WD. $L_{\mathrm{Edd}}$ is the Eddington luminosity of the RGB core at the moment when the CE is initiated and, assuming accretion processes on the core by the companion, is the maximum possible luminosity that can be achieved. 
Both the values are taken from the {\sc mesa} model that has been mapped into {\sc phantom} for the hydrodynamic simulation and correspond to $L_{\mathrm{core}} \simeq 10^{2.81}$~\ls \ and $L_{\mathrm{Edd}} \simeq 10^{4.32}$~\ls.

The dependence of $T_{\mathrm{ph}}$ on the the homologous expansion kinematics and on the local properties of the medium comes into play when determining the radius of the photosphere. $R_{\mathrm{ph}}$ can be obtained by numerically integrating the optical depth inwards, $\tau = \int \kappa \rho dR$, where $\kappa$ is the opacity (more accurately, $\kappa$ is the mass absorption coefficient of dust, nomenclature we will adopt instead of using the word \virg opacity''). $R_{\mathrm{ph}}$ is then obtained as the radius where the gas becomes optically thick ($\tau = 2/3$). Assuming homologous expansion, the optical depth equation can be rewritten as
\begin{equation}
\tau = \Big( \frac{t}{t_{\rm i}} \Big)^{-2} \int \kappa \rho_{\rm i} dr_{\rm i} \ ,
 \label{eq:homologous_optical_depth}
\end{equation}
where $t_{\rm i}$, $\rho_{\rm i}$ and $r_{\rm i}$ are time, density and pressure at the moment when homologous expansion kicks in. In our case at 5000~days. 
In this way the optical depth is a function of both the optical properties of the gas and of the initial homologous quantities, as a result $R_{\mathrm{ph}}$ depends on the same quantities.

For the low temperature situation considered here, we assume that the mass absorption coefficient $\kappa$ is dominated by newly formed dust within the CE ejecta. In the accompanying paper, we will justify this assumption by showing that the unbound CE ejecta satisfy the conditions for dust formation after 5000~days (Iaconi et al., in preparation; Reichardt et al., in preparation). Similarly to the case of dust formation in supernova ejecta, the main contributions to the mass absorption coefficient may well come from MgSiO$_3$ (silicates) or carbon dust (\citealt{Nozawa2008}, \citealt{Maeda2013}). We will therefore ignore any other source of opacity and focus on the MgSiO$_3$ and carbon dust grains. We will analyse two possible regimes: dust component dominated by MgSiO$_3$ and dust component dominated by carbon.
RGB star chemistry is dominated by oxygen-based dust because the gas phase C/O ratio is lower than unity. However, we experiment with the case of carbon dust because AGB stars, in many ways similar to RGB stars, can have C/O ratios above unity and be dominated by carbon type dust. Using both types will therefore be informative for future work.

We first verify that dust could actually form and remain stable in the expanding envelope by using the procedure of \citet{Nozawa2013} and dust parameters from \citet{Nozawa2003}. We find that, at $\simeq 300$~days, when the dynamic in-spiral terminates, the some SPH particles cool down enough that dust can form at the outskirts of the expanding envelope. After that, dust keeps forming until $\simeq 5000$~days. The dust that forms is stable according to the criterion from \citet{Nozawa2013} we used here. At the time when homologous expansion sets in, dust has therefore formed over the entire envelope. This ensures that the entire amount of dust that can possibly form, actually forms, both in the case of MgSiO$_3$ and carbon. Details of the dust formation processes and calculations will be presented elsewhere (Iaconi et al., in preparation). Assuming that all the dust-forming metals are confined either to MgSiO$_3$ or carbon grains, and using a Solar composition, the amount of dust in the whole ejecta is $\simeq 2.9 \times 10^{-4}$ (0.06 percent of the envelope mass) and $\simeq 2.5 \times 10^{-3}$~\ms \ (0.5 percent) in the former and latter cases, respectively.

All the quantities at the photosphere location for the range of dust grain sizes and as a function of time are determined by sampling the density distribution at 5000~days into a series of concentric radial shells (a total of $10^6$ shells), numerically performing the integral in Equation~\ref{eq:homologous_optical_depth} on the initial distribution and by evolving over time the value of $\tau$ according to its dependence on the homologous factor $t/t_{\rm i}$ up to 18\,250~days (for a total of $10^4$ steps). The mass absorption coefficients necessary for the calculations are taken from the tables used by \citet{Nozawa2008}, which report mass absorption coefficient as a function of wavelength for four different dust grain sizes (0.001$\mathrm{\mu m}$, 0.01~$\mathrm{\mu m}$, 0.1~$\mathrm{\mu m}$, 1~$\mathrm{\mu m}$).
For each time-step, we compute all the possible $T_{\mathrm{ph}}$ from all the wavelengths available in the tables through the process described above. The (wavelength-averaged) photospheric temperature is then determined by using the Wien's displacement law, which provides a relation between $T_{\mathrm{ph}}$ and the peak wavelength of the black-body.

\subsection{Location and temperature of the photosphere}
\label{ssec:location_and_temperature_of_the_photosphere}
\begin{figure*}
\centering
\subfigure[]{\includegraphics[scale=0.45, trim=0.0cm 0.0cm 0.0cm 0.0cm]{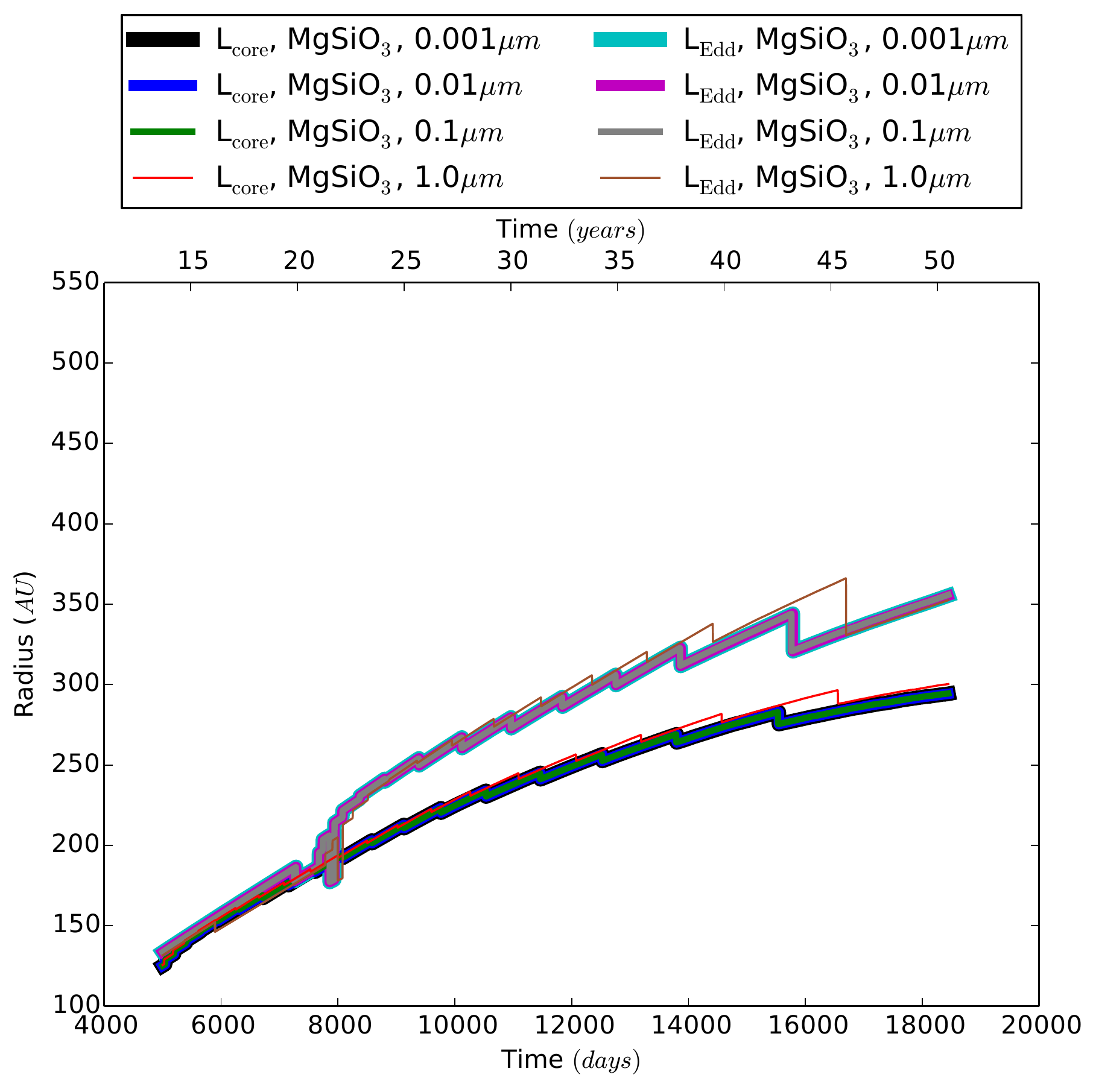}}
\subfigure[]{\includegraphics[scale=0.45, trim=0.0cm 0.0cm 0.0cm 0.0cm]{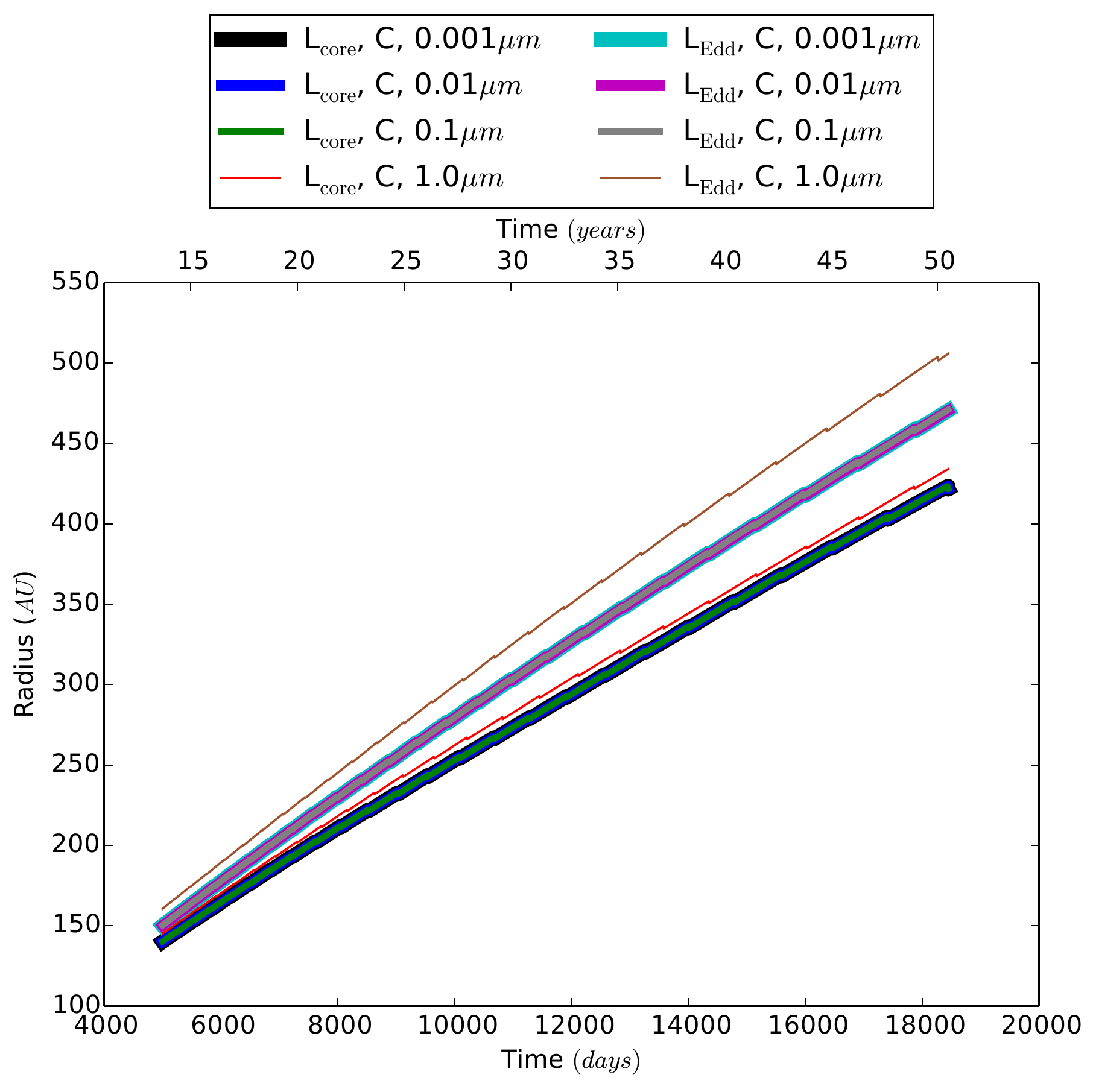}}
\caption{\protect\footnotesize{Location of the photosphere as a function of time for two different dust types: MgSiO$_3$ (a) and carbon (b). The calculations have been performed for L$_{\mathrm{core}}$ and L$_{\mathrm{Edd}}$ and for four dust grain sizes, as described in the legend.}}
\label{fig:photosphere_radius_vs_time}
\end{figure*}

\begin{figure*}
\centering
\subfigure[]{\includegraphics[scale=0.45, trim=0.0cm 0.0cm 0.0cm 0.0cm]{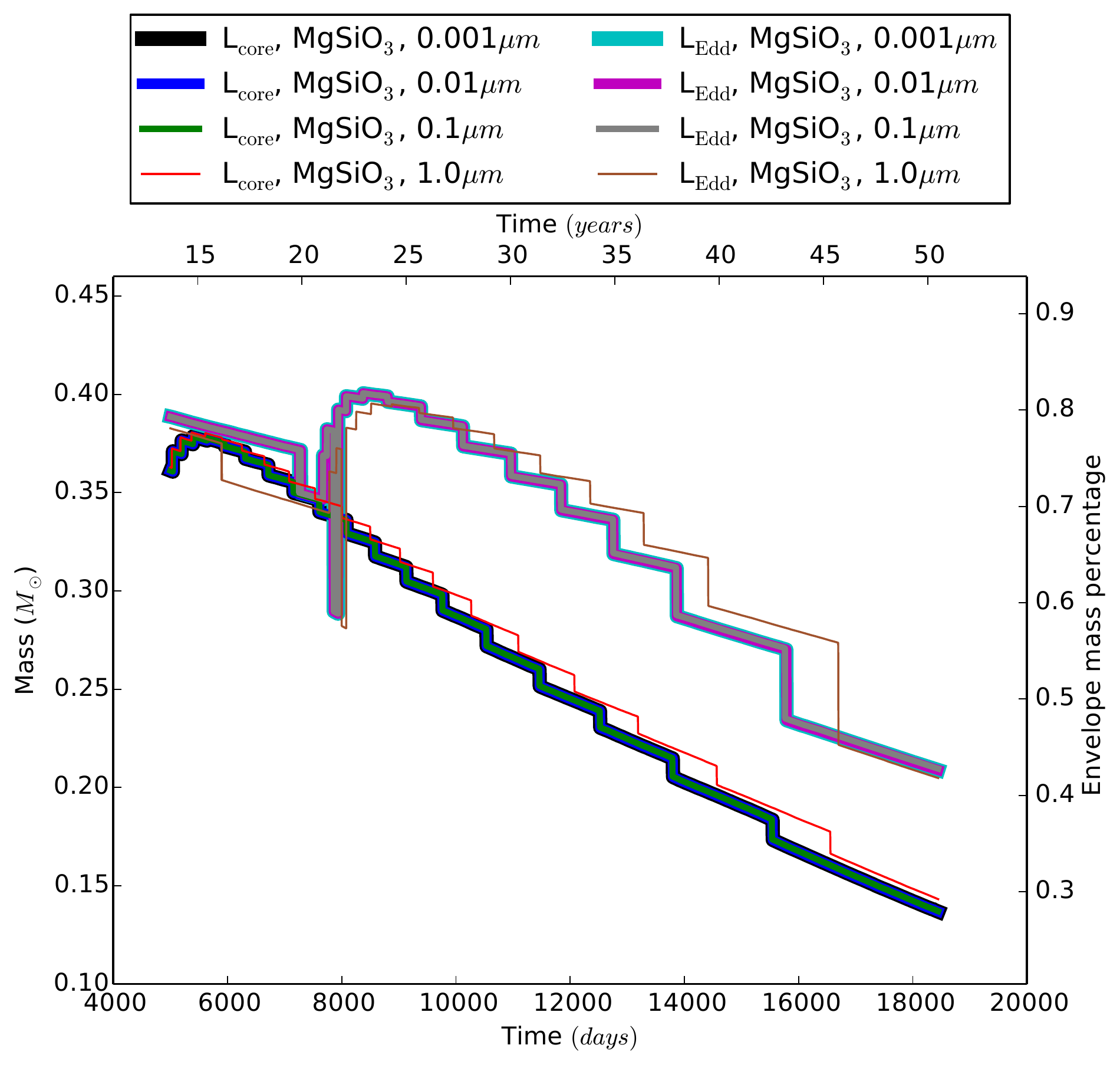}}
\subfigure[]{\includegraphics[scale=0.45, trim=0.0cm 0.0cm 0.0cm 0.0cm]{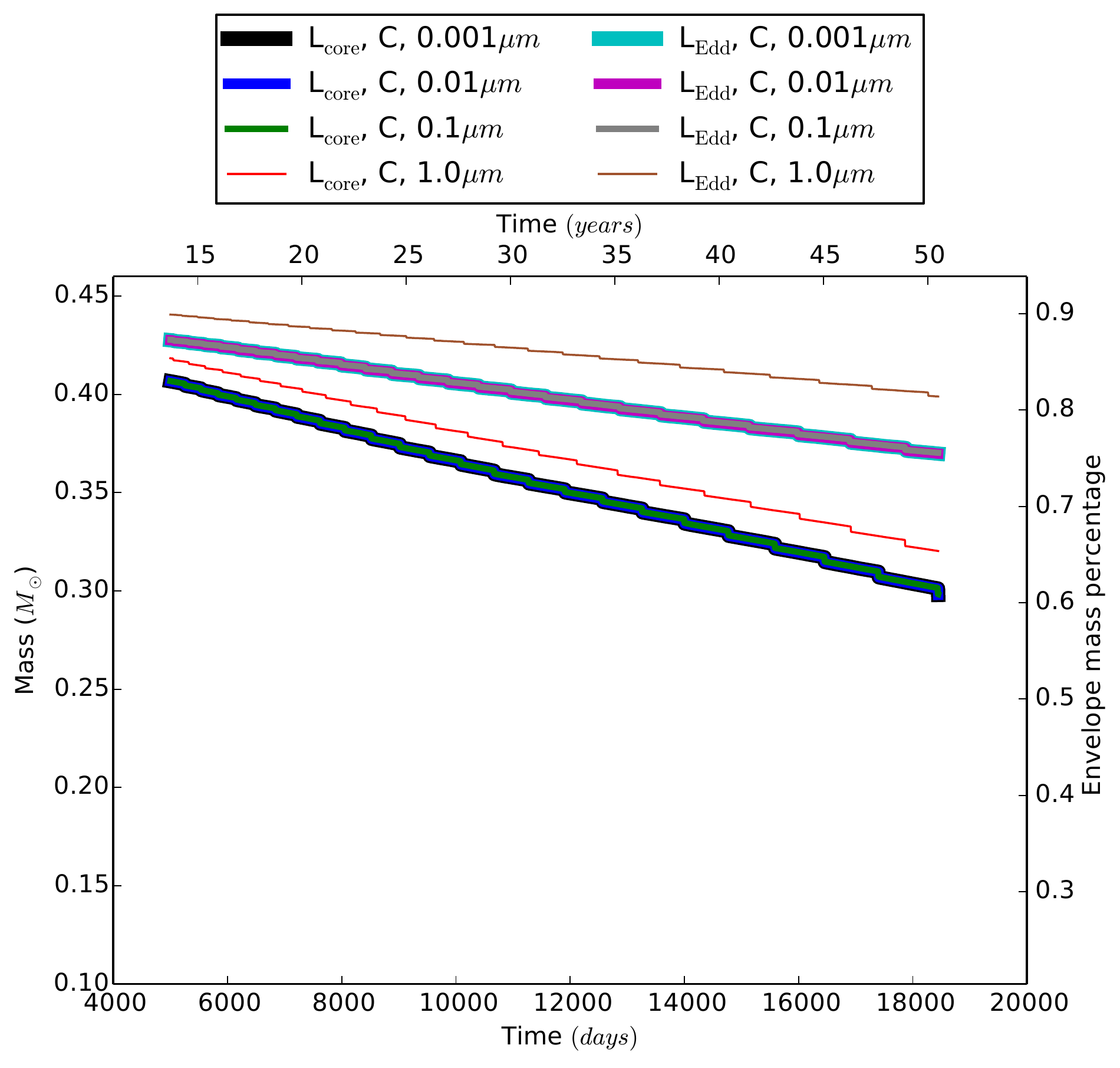}}
\caption{\protect\footnotesize{Envelope mass enclosed in the photosphere as a function of time for two different dust types: MgSiO$_3$ (a) and carbon (b). The calculations have been performed for L$_{\mathrm{core}}$ and L$_{\mathrm{Edd}}$ and for four dust grain sizes, as described in the legend.}}
\label{fig:photosphere_mass_vs_time}
\end{figure*}

We show the estimated location of the photosphere in Figure~\ref{fig:photosphere_radius_vs_time}, while in Figure~\ref{fig:photosphere_mass_vs_time} we plot the envelope mass enclosed in the photosphere as a function of time.
In both cases the zig-zag appearance of the curves is a numerical artefact due to the simplified treatment of the temperature determination coupled with the wavelength discretisation in the mass absorption coefficient tables.

The difference between the behaviour of MgSiO$_3$ and carbon is easily noticeable and depends on how the dust absorption coefficients of the two dust types depend on wavelength, the former showing a more complex evolution as a function of time. 
The location of the photosphere is at larger radii for the carbon dust with respect to MgSiO$_3$. This depends in part on the optical properties of the carbon dust and in part on the abundance of carbon in the envelope gas. 

If we compare the radii achieved by the particles' distribution (Figure~\ref{fig:radial_velocities_multiple_timestep}, last two panels) with the photospheric radii for the two dust types, we can see that in both cases the location of the photosphere is inside the region where the bulk of the SPH particles are. The gas residing in the external tail of the radial distribution has an extremely low density and generates a minimally contributes to the total optical depth. It is interesting to see that the photosphere is not at a fixed homologous coordinate, i.e., even though its location increases over time, it falls behind the homologous expansion velocity and moves gradually inwards. This is especially true in the case of MgSiO$_3$ dust, for which the increase in the photospheric radius becomes less steep as time passes. As a result of this the mass enclosed in the photosphere decreases over time (Figure~\ref{fig:photosphere_mass_vs_time}). In the case of MgSiO$_3$ the initial mass enclosed in the photospheric radius is smaller than for carbon and the percentage decrease over time is larger for silicates than carbon. 

\begin{figure*}
\centering
\subfigure[]{\includegraphics[scale=0.45, trim=0.0cm 0.0cm 0.0cm 0.0cm]{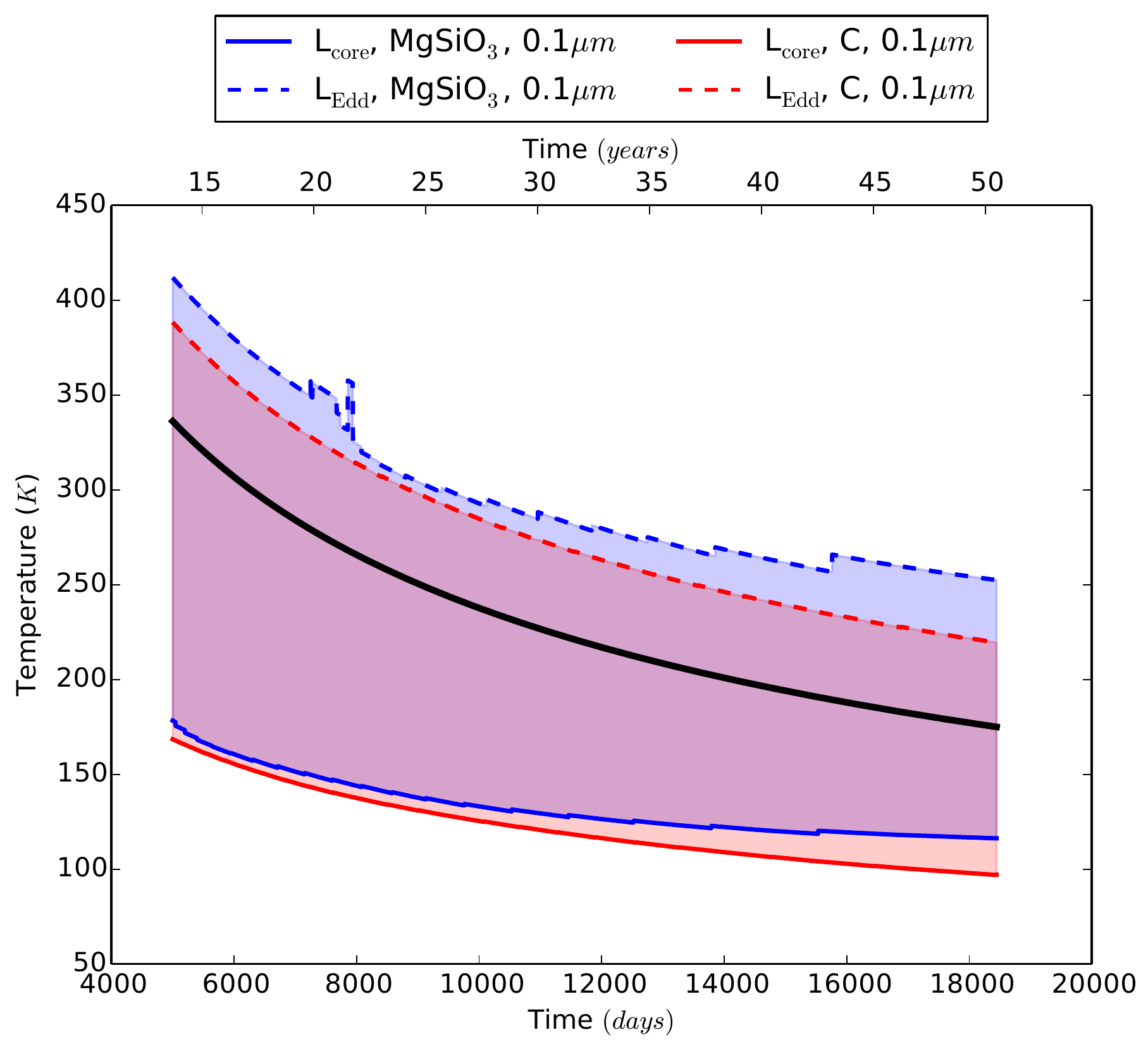}}
\subfigure[]{\includegraphics[scale=0.45, trim=0.0cm 0.0cm 0.0cm 0.0cm]{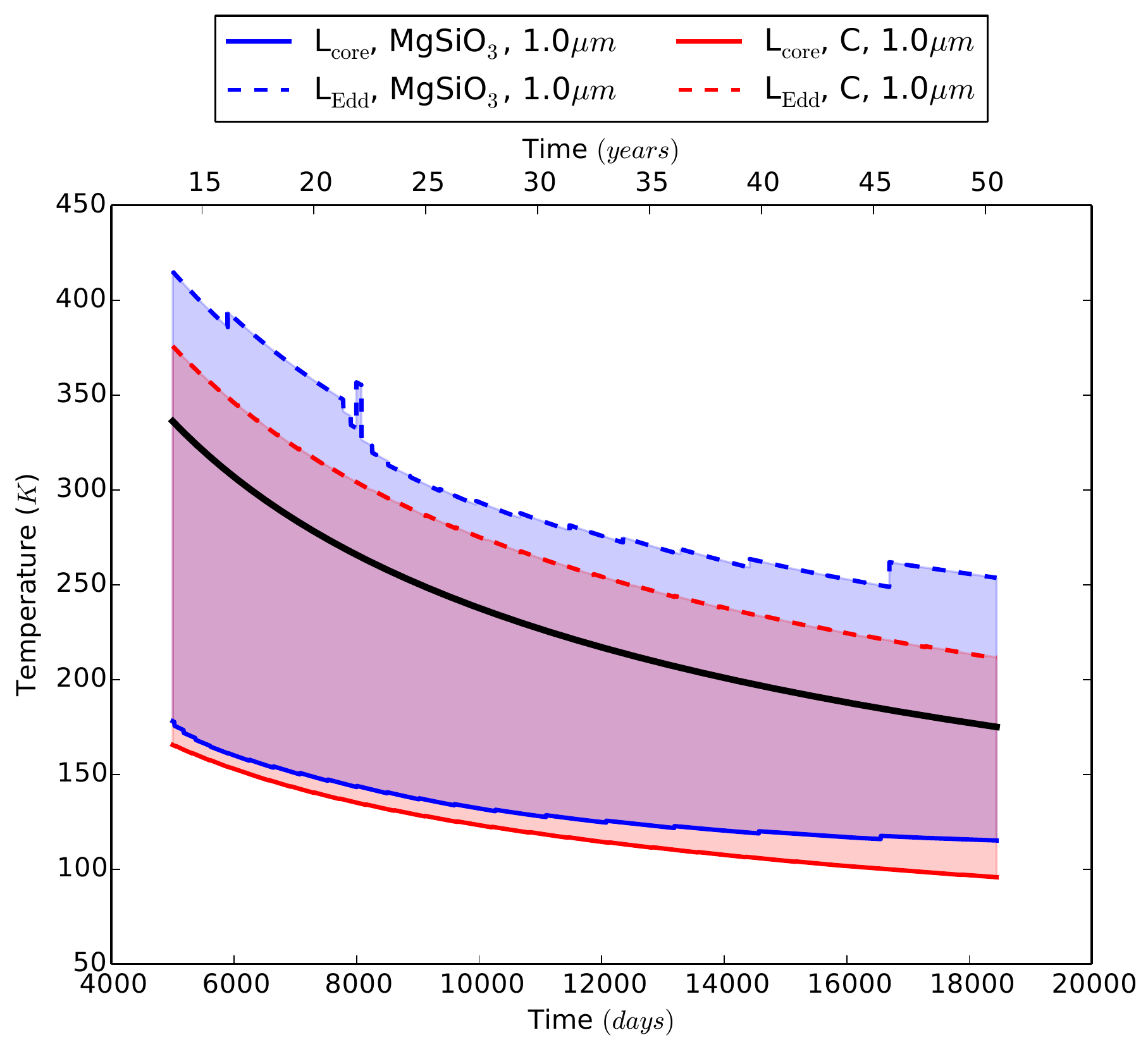}}
\caption{\protect\footnotesize{Temperature of the photosphere as a function of time plotted for two different dust grain sizes (0.1$\mathrm{\mu m}$ (a) and 1~$\mathrm{\mu m}$ (b)). As specified in the text, the remaining grain sizes show a behaviour exactly identical to that of panel (a), therefore have not been plotted here. For each grain size MgSiO$_3$ dust and C dust at both L$_{\mathrm{core}}$ and L$_{\mathrm{Edd}}$ are represented. To highlight the range of possible temperatures we shaded the respective areas between L$_{\mathrm{core}}$ and L$_{\mathrm{Edd}}$. The black think line is a $t^{-\frac{1}{2}}$ curve draw for comparison.}}
\label{fig:tempphoto_radius_vs_time}
\end{figure*}

In Figure~\ref{fig:tempphoto_radius_vs_time} we show the temperature at the photosphere location for the 0.1$\mathrm{\mu m}$ and 1~$\mathrm{\mu m}$ dust grains. We do not plot the results for the 0.01$\mathrm{\mu m}$ and 0.001~$\mathrm{\mu m}$ dust grains because their behaviour is identical to the 0.1$\mathrm{\mu m}$ case. This is because for the range of temperatures we consider ($\simeq 100 - 400$~K), the mass absorption coefficient behaves similarly for the 0.001$\mathrm{\mu m}$, 0.01$\mathrm{\mu m}$ and 0.1$\mathrm{\mu m}$ grains and as a result also the optical depth is independent of the grain radius.

The receding photosphere discussed in the context of Figure~\ref{fig:photosphere_mass_vs_time} is also apparent in Figure~\ref{fig:tempphoto_radius_vs_time} by observing that the temperature evolves with a shallower slope with respect to that expected from Equation~\ref{eq:stefan_boltzmann_law} and homologous expansion, namely $T_{\mathrm{ph}} \propto t^{-\frac{1}{2}}$ (black line in Figure~\ref{fig:tempphoto_radius_vs_time}). The $T_{\mathrm{ph}} \propto t^{-\frac{1}{2}}$ evolution assumes that the photosphere does not move in homologous coordinates. This would be true if $\tau \gg 2/3$ with the photosphere residing at the outer edge of the mass distribution. However, the location of $\tau = 2/3$ resides inside the envelope and, as the density decreases during the expansion, the photosphere automatically starts to move inward. This causes the trend we observe in Figure~\ref{fig:photosphere_radius_vs_time}, \ref{fig:photosphere_mass_vs_time} and \ref{fig:tempphoto_radius_vs_time}.

With this in mind, we observe that for all the grain sizes the combination of carbon dust and a central luminosity of L$_{\mathrm{Edd}}$ yields the temperature evolution closest to the homologous evolution. This also corresponds to the case where the mass enclosed inside the photosphere decreases by the smallest amount. In the other cases, since the photosphere moves closer to the central binary, the temperature shows a flatter decrease.

If we observe the behaviour of the density (Figure~\ref{fig:density_vs_radius}), we notice that at 9125~days, the density at the location of the photosphere tends to be flatter inside the silicates dust photosphere with respect to the photosphere obtained for carbon dust, this supports the difference in recession we observe between silicates and carbon (i.e., the photosphere in the case of silicates dust moves inward faster than in the case of carbon dust). The photosphere is more inside for silicates than carbon dust and if the density distribution is flat then the speed of the photosphere recession is higher at smaller radii.

\begin{figure*}
\centering
\subfigure[]{\includegraphics[scale=0.45, trim=0.0cm 0.0cm 0.0cm 0.0cm]{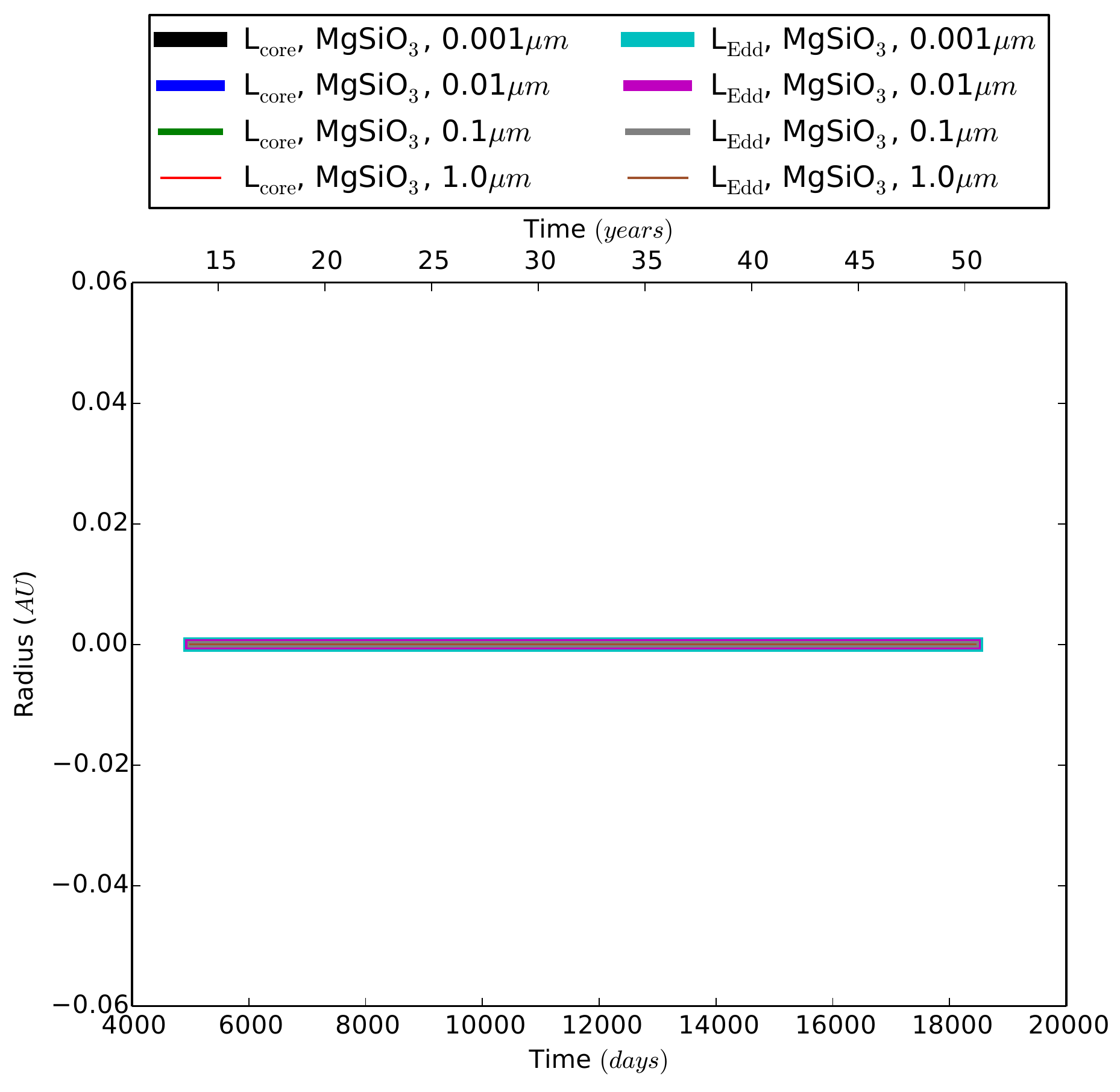}}
\subfigure[]{\includegraphics[scale=0.45, trim=0.0cm 0.0cm 0.0cm 0.0cm]{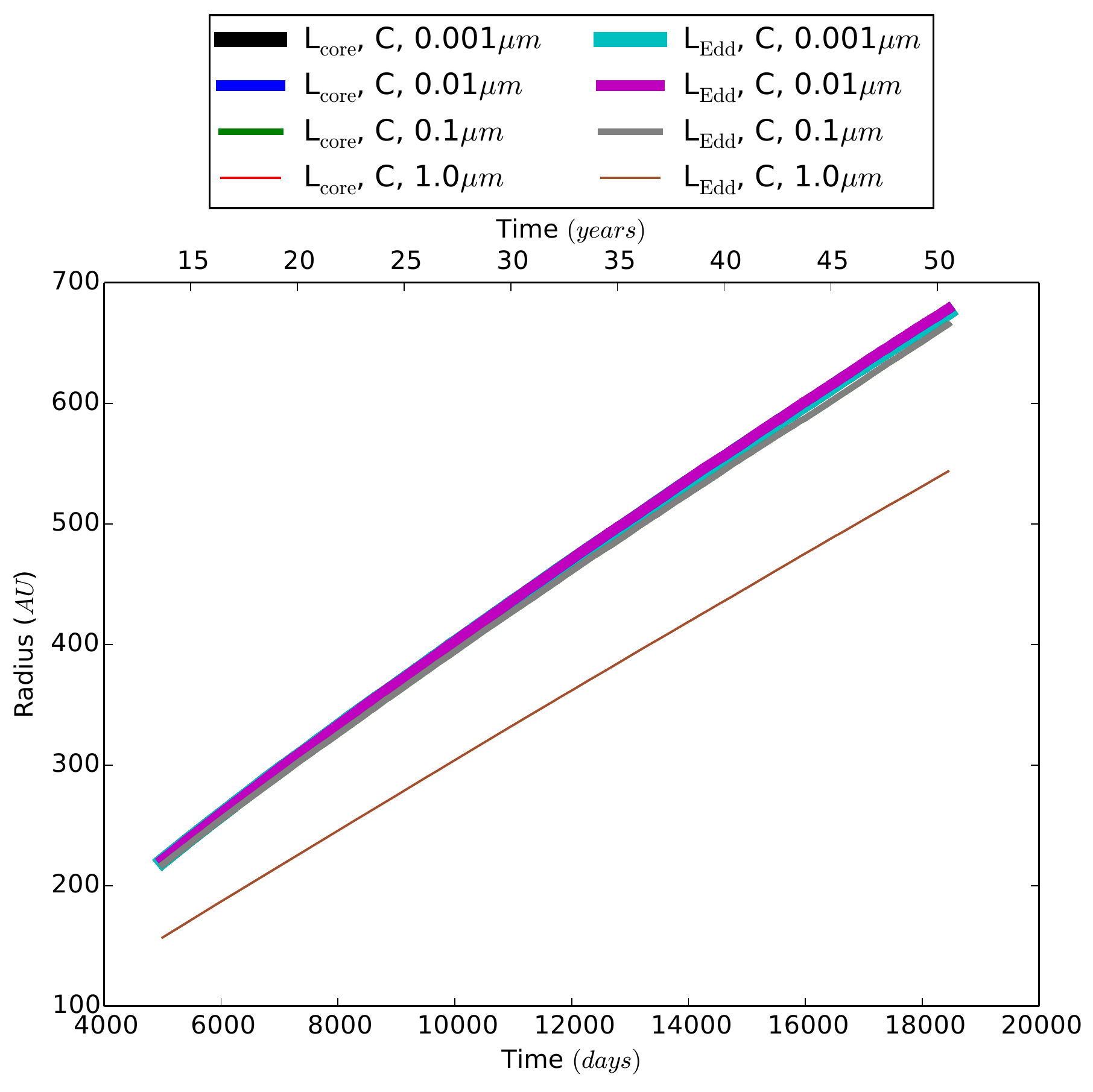}}
\caption{\protect\footnotesize{Same as Figure~\ref{fig:photosphere_radius_vs_time}, but for a fixed wavelength of $\simeq 0.4$~$\mu m$.}}
\label{fig:photosphere_radius_vs_time_0d4um}
\end{figure*}

\begin{figure*}
\centering
\subfigure[]{\includegraphics[scale=0.45, trim=0.0cm 0.0cm 0.0cm 0.0cm]{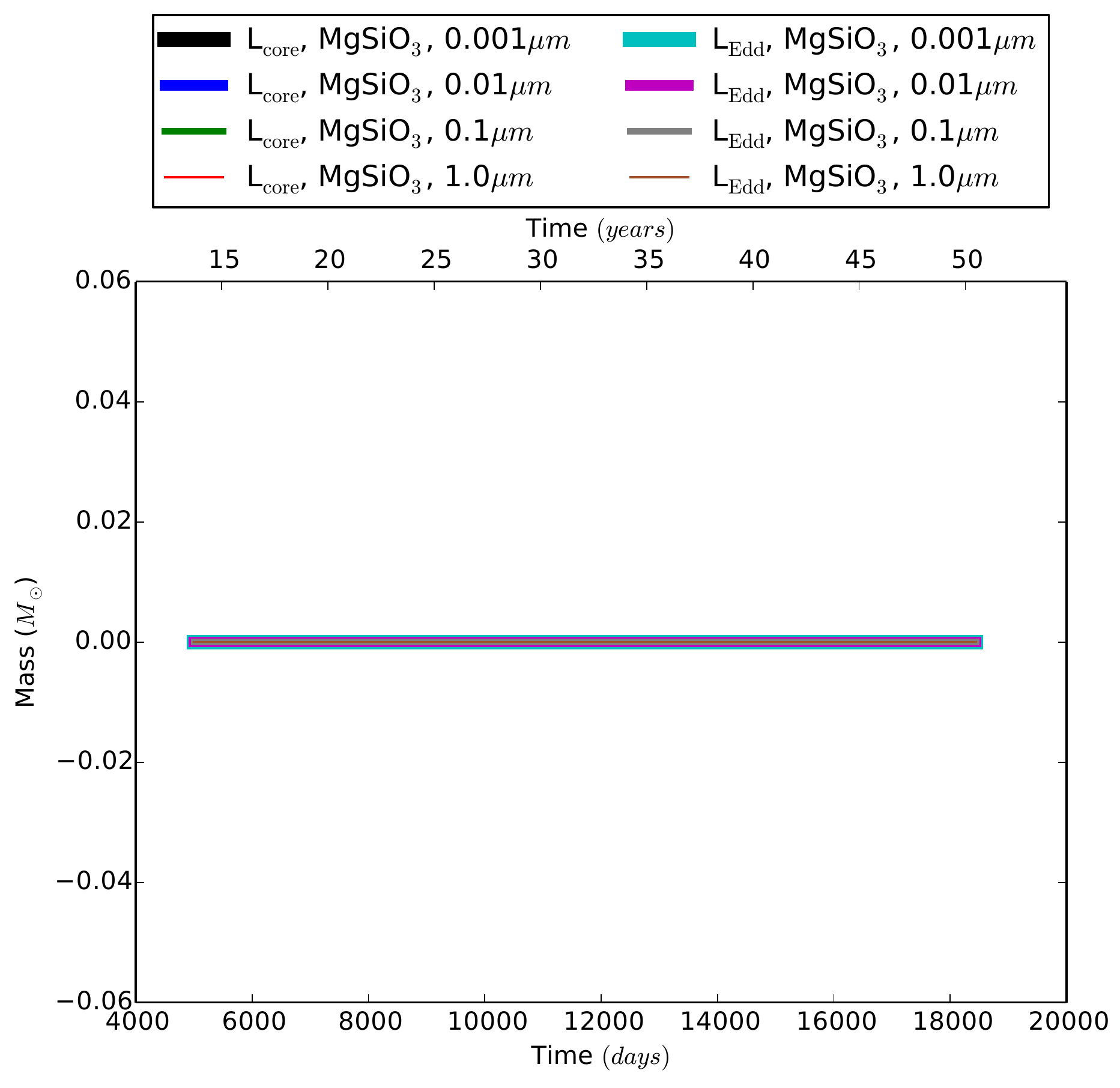}}
\subfigure[]{\includegraphics[scale=0.45, trim=0.0cm 0.0cm 0.0cm 0.0cm]{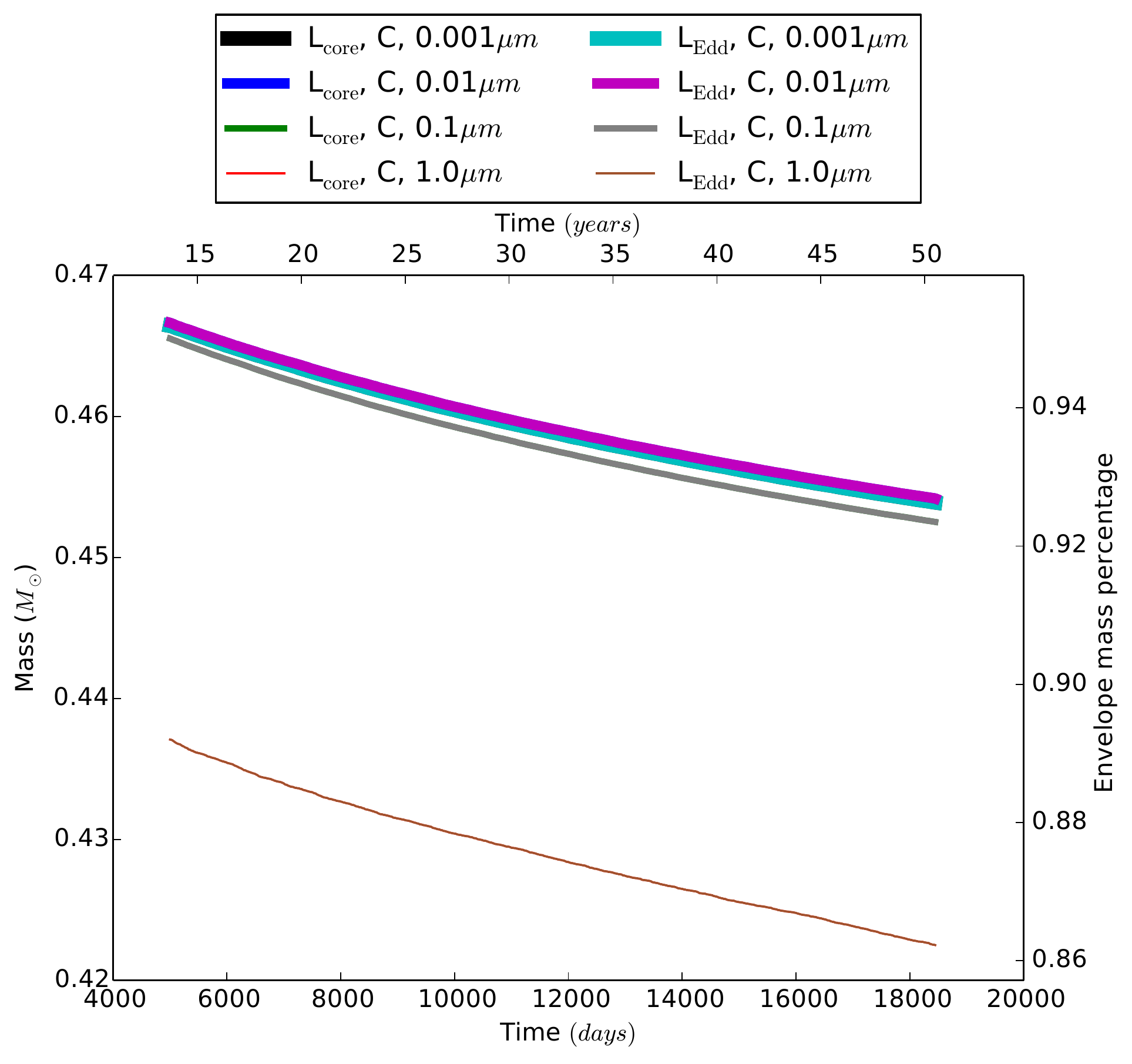}}
\caption{\protect\footnotesize{Same as Figure~\ref{fig:photosphere_mass_vs_time}, but for a fixed wavelength of $\simeq 0.4$~$\mu m$.}}
\label{fig:photosphere_mass_vs_time_0d4um}
\end{figure*}

Typical photospheric temperatures correspond to black-body emission in the IR band. However, the dust mass absorption coefficient is wavelength dependent and so is the optical depth. As a sanity check, we performed the same integration at a fixed wavelength of 0.4~$\mu m$, in the optical band. At this wavelength, MgSiO$_3$ has a very low mass absorption coefficient, while the carbon one is very high. Indeed this results in an optically thin envelope in the former case and in a optically thick one in the latter (Figures~\ref{fig:photosphere_radius_vs_time_0d4um} and \ref{fig:photosphere_mass_vs_time_0d4um}). In the carbon case the location of the photosphere is in line with the various mass absorption coefficients at different grain sizes reported in the tables. This confirms that our procedure of integration behaves correctly. On the other hand, in case of MgSiO$_3$, only a fraction of the luminosity from the central system may indeed be absorbed within the ejecta, which may reduce the expected temperature of the emission component associated with the ejecta.

Our estimate of the photospheric temperature is not very precise, and a more detailed study would require a full radiation transfer calculations. However, based on the geometry of the ejecta we can predict features that may appear in post dynamic in-spiral observations. In the first few years after the dynamic in-spiral the ejected material has a larger density on the orbital plane than in other directions. Therefore we expect dust obscuration primarily along the orbital plane, while we would see radiation from a more central location or even from the binary itself when observing the system face-on. In this situation the IR emission from the dusty torus would be in addition to that emitted by the central object and result in a IR excess. This is essentially a IR echo by the optically thick material added to the unobscured optical/NIR light from the central system. As time passes the envelope distribution becomes gradually more symmetric and the shells discussed in Section~\ref{ssec:cocoon_formation} form. We can therefore expect that the IR excess would gradually die out and that the system would look similar when observed from different angles, once it has achieved symmetry.

There are already several observations of transients, thought to derive from common envelope mergers, that developed the signature of newly formed dust. Examples are V~1309~Sco that displayed a 1-magnitude dip just before the outburst that marked the moment of the merger \citep{Nicholls2013}, V~838~Mon \citep{Bond2003} and more recently the luminous red nova AT~2017jfs in NGC~4470, which is a massive counterpart of the first two \citep{Pastorello2019}. Ultimately surveys such as the {\it Spitzer} survey SPIRITS \citep{Kasliwal2017} and telescopes such as the dedicated Palomar Gattini-IR \citep{Moore2016} will allow us to discover the IR signature of common envelope transients due to the dust that forms in the expanding envelope. 

\section{Summary and conclusions}
\label{sec:summary_and_conclusions}
In this work we analysed the behaviour of the ejecta from a CE interaction simulated with the 3D SPH hydrodynamic code {\sc phantom} over a post dynamic in-spiral time-scale much longer than usually achieved. We carry out a CE simulation with the same setup used by \citet{Passy2012}, namely, a primary RGB star with an initial mass of 0.88~\ms \ and an initial radius of 83~\rs, plus a point-mass companion of 0.6~\ms. The companion is placed on the primary surface in circular orbit at the beginning of the simulation. Additionally, we utilise a tabulated equation of state that allows us to include the recombination energy into the gas and achieve a full unbinding of the envelope.

During the long post-in-spiral time-scale the ejecta have time to reach a regime where they evolve without any injection or loss of energy.
We propose that, since this gas does not receive any additional energy and is unbound, the evolution of the ejecta can be modelled as a homologously expanding ideal gas under adiabatic conditions.\\
Our main results are the following:

\begin{enumerate}
    \item A key condition for the homologous expansion model to represent the data is that the velocity field must be dominated by the radial component of the velocities. We observe that during the dynamic in-spiral (which lasts $\simeq 300$~days) and until $\simeq 1800$~days from the beginning of the simulation, the tangential velocities remain non-negligible. However, after $\simeq 1800$~days the fraction of gas with high radial velocities rapidly increases and dominates the gas distribution. Because of this we can apply a homologous model to the numerical data.
    
    \item Since the energy injection time-scale in CE interactions is not instantaneous as is the case in supernovae, envelope layers ejected at different times tend to be best approximated by different choices for the initial homologous time, $t_0$. This is mostly evident for the envelope layers ejected early in the dynamic in-spiral, whose distribution rapidly converges towards the analytical relation in the radial velocity vs. radius plane. We find that if $t_0$ is chosen during the first $\simeq 620$~days from the beginning of the simulation, time during which all the possible energy is injected into the envelope, any choice of $t_0$ becomes a viable approximation to describe the homologous expansion of the envelope. This becomes evident once the average thermodynamic quantities of the envelope start evolving in accordance to the homologous regime ($\simeq 5000$~days, more on this in point \ref{item_four}).
    
    \item When the system achieves homologous expansion, we observe that morphology and density profiles do not change shape as time passes, in line with the behaviour expected from a homologously expanding system.
    The result of this behaviour is the formation of a series of expanding shells with decreasing average densities from the central binary (Figure~\ref{fig:asymptotic_slices}, bottom panels). Since the gas is expanding homologously, the shells tend to move apart from each other as time passes, with the external shells moving faster than the internal ones.

    \item Thermal energy, temperature, pressure, density and entropy evolution initially deviate from the homologous evolution as the dynamic in-spiral and energy injection by recombination take place (in the first $\simeq 620$~days of the simulation). The difference persists until $\simeq 5000$~days, during which the bulk of the ejecta are still carrying the residuals of the previous, turbulent interaction. After this the chaotic gas distribution becomes more ordered.
    After $\simeq 5000$~days the main thermodynamic quantities move towards their homologous values. Eventually all quantities evolve homologously until the end of the simulation, showing therefore a decrease as a power-law of time for the main thermodynamic quantities ($E_{\mathrm{therm}} \propto t^{-2}$, $T \propto t^{-2}$, $P \propto t^{-5}$, $\rho \propto t^{-3}$ and $S \propto \rm{const.}$). This shows that the time-scales for the envelope to become homologous are a factor of 10 longer than those of the dynamic in-spiral.\label{item_four}

    \item Utilising the homologous model proposed here and the dust formation model by \citet{Nozawa2013}, we calculate the location of the photosphere on the orbital plane of the ejecta starting at 5000~days after the beginning of the in-spiral. We find that,  irrespective of the dust type used (silicates and carbon dust) and of the dust grain size, the photosphere is always located within the outer edge of the dense part of the ejecta, where the bulk of the envelope mass resides. The photosphere does not expand homologously, but slowly moves inwards. This happens faster for silicates than for carbon dust. From 5000~days onward the photosphere has temperatures between $\simeq 100$~K and $\simeq 450$~K, compatible with emission in the IR band. We also highlight that the behaviour of the photosphere does not really rely on the homologoues expansion, and qualitatively a similar behaviour is expected even in the phases where the homologous expansion does not apply. 
\end{enumerate}

Concluding, we have studied a part of the CE evolution that had not been considered previously in detail. 
Interestingly, a complex phenomenon like the CE dynamic in-spiral can be described with a rather simple analytic model in its asymptotic regime. We here show how this simplifies the determination of the photosphere location and temperature. A similar approach can be applied alongside a full radiation transfer calculations, as is done for supernovae. 
This would allow us to calculate a synthetic light-curve for the post dynamic in-spiral ejecta to be compared with observations of transients, bypassing many of the obstacles encountered by the approach of \citet{Galaviz2017}.


\section*{Acknowledgments}
\label{sec:acknowledgments}
RI is grateful for the financial support provided by the Postodoctoral Research Fellowship of the Japan Society for the Promotion of Science (JSPS P18753) and by the International Macquarie University Research Excellence Scholarship. KM acknowledges the support provided by the JSPS KAKENHI grant 18H04585, 18H05223, and 17H02864. TN acknowledges the support provided by the JSPS KAKENHI grant 18K03707.
We are grateful to the referee, Noam Soker, for the useful comments and for helping to improve the paper. We also thank Brian Metzger and Tomasz Kaminski for their feedback after posting this work on astro-ph.
We acknowledge the computational facilities of the Department of Physics and Astronomy of Macquarie University, on which the simulations have been carried out. The computations described in this work were performed using the PHANTOM code (https://phantomsph.bitbucket.io/), which is the product of a collaborative effort of scientists under the lead of A/Prof. Daniel Price (Monash University, Melbourne, VIC, Australia).
All simulation outputs are available upon request by e-mailing roberto.iaconi@kusastro.kyoto-u.ac.jp.


\bibliographystyle{aa}
\bibliography{bibliography}{}
\bsp


\label{lastpage}
\end{document}